\documentstyle[12pt]{article}
\newcommand{\bce}{\begin{center}}
\newcommand{\ece}{\end{center}}
\newcommand{\beq}{\begin{equation}}
\newcommand{\eeq}{\end{equation}}
\newcommand{\bea}{\vspace{0.25cm}\begin{eqnarray}}
\newcommand{\eea}{\end{eqnarray}}

\newcommand{\ba}{\begin{array}}
\newcommand{\ea}{\end{array}}

\newcommand{\doublespace}{
    \renewcommand{\baselinestretch}{1.6}\large\normalsize}

\def\lsim{\mathrel{\rlap{\lower4pt\hbox{\hskip1pt$\sim$}}
    \raise1pt\hbox{$<$}}}     
\def\gsim{\mathrel{\rlap{\lower4pt\hbox{\hskip1pt$\sim$}}
    \raise1pt\hbox{$>$}}}     

\def\lsim{\mathrel{\rlap{\lower4pt\hbox{\hskip1pt$\sim$}}
    \raise1pt\hbox{$<$}}}         
\def\gsim{\mathrel{\rlap{\lower4pt\hbox{\hskip1pt$\sim$}}
    \raise1pt\hbox{$>$}}}         

\def\beq{\begin{equation}}
\def\endeq{\end{equation}}
\def\arr{\begin{eqnarray}}
\def\endarr{\end{eqnarray}}

\makeindex
\begin{document}
\vspace{2cm}
\begin{center}
{\bf \huge Do final state interactions obscure
short range correlation effects
in quasielastic
$A(e,e'p)$ scattering ? \\}
\vspace{1cm}
{\bf A.Bianconi$^{1)}$, S.Jeschonnek$^{2)}$,
N.N.Nikolaev$^{2,3)}$, B.G.Zakharov$^{3)}$ } \medskip\\
{\small \sl
$^{1)}$Dipartimento di Fisica Nucleare e Teorica, Universit\`a
di Pavia, and\\
Istituto Nazionale di Fisica Nucleare,
Sezione di Pavia, Pavia, Italy \\
$^{2)}$IKP(Theorie), Forschungszentrum  J\"ulich GmbH.,\\
D-52425 J\"ulich, Germany \\
$^{3)}$L.D.Landau Institute for Theoretical Physics, \\
GSP-1, 117940, ul.Kosygina 2, V-334 Moscow, Russia
\vspace{1cm}\\}
{\bf \LARGE A b s t r a c t \bigskip\\}
\end{center}
Are short range correlations in the ground state of the target
nucleus (initial state correlations ISC)
observable in experiments on quasielastic $A(e,e'p)$
scattering at large missing momentum $p_{m}$? Will the
missing momentum spectrum observed at CEBAF be overwhelmed
by final state interactions of the struck proton? Taking the
$^{4}He$ nucleus with a realistic model wave function
for a testing ground,
we present a full calculation of the missing
momentum distribution in inclusive $^{4}He(e,e'p)$ scattering.
We find a complex interplay and strong quantum-mechanical
interference of FSI and ISC
contributions to scattering at large $p_{m}$, with drastic change
of the interference pattern from the (anti)parallel to
transverse kinematics.
We show that in all the kinematical
conditions, for missing momenta $p_{m}\gsim 1$\,fm$^{-1}$,
quasielastic scattering is dominated by FSI effects and
the sensitivity to details of the nuclear
ground state is lost.  The origin of the FSI dominance is well
understood and can be traced back to the anisotropic behaviour of FSI
which is long ranged in the longitudinal direction and short ranged in
the transverse direction
in the opposite to the short ranged
 ground state correlations.

\newpage
\doublespace

\section{Introduction}

Investigation of short-distance nucleon-nucleon interaction
(initial state two-nucleon correlations (ISC)) in the nuclear
medium is considered one of the principal goals
of experiments on quasielastic $A(e,e'p)$ scattering at large
missing momentum $p_{m}$ \cite{reviews}.
Although the principal ideas and
motivations for such experiments
go back to Gottfried's classic works of the early 60's
(\cite{Gottfried}, see also \cite{SrivYu,Banerjee}), they
are only becoming feasible at a new generation of high
luminosity, continuous beam, electron facilities (CEBAF, AmPS,
MAMI, Bates). Of special importance is a new domain of
large $(e,e')$ momentum transfer squared $Q^{2}$ attainable at
CEBAF [5], because for the first time the
kinetic energy $T_{kin}\approx Q^{2}/2m_{p}$ of the struck
proton will be high enough to exhaust the missing energy spectrum.

The experimentally measured $p_{m}$ distribution is distorted
by final state interaction (FSI) of the struck proton in the
target nucleus debris. Are these final state interactions strong
or do they just lead to small corrections to the ISC contribution to
large-$p_{m}$ phenomena? The answer to this pressing question
and the mere possibility of the theoretical interpretation of
the forthcoming CEBAF high-$Q^{2}$ experimental data on large $p_{m}$
in terms of the ground state correlation
effects requires the quantitative understanding of FSI effects.
Several aspects of FSI in the high energy regime of
CEBAF experiments have already been discussed in
\cite{Pperp,NE18CT,NNZ,Helium4,Deuter,NSZ,Tensor} with
the important and disturbing finding
that FSI effects completely take over and
make the large-$p_{m}$
behaviour of the experimentally observed
missing momentum distribution drastically
different from the single-particle momentum distribution (SPMD)
in the ground state of the target nucleus. FSI effects were found
to be strong in even such a dilute nucleus as the deuteron
\cite{Deuter,Tensor}.
It is obvious that at large $Q^{2}\gsim (1-2)\,GeV^{2}$ and large
$T_{kin}\approx Q^{2}/2m_{p}$, the conventional potential model
description
of FSI becomes impractical.
The major point is that the
very nature of nucleon-nucleon scattering changes from the
purely elastic,
potential scattering at low energies to a
strongly absorptive, diffractive small angle scattering at $T_{kin}
\gsim
(0.5-1)\,GeV$. In this diffractive  regime,
Glauber's multiple scattering theory \cite{Glauber} becomes a
natural framework for quantitative description of FSI and leads
to several important new effects in the calculation of FSI-modified
one-body density matrix and missing momentum distribution in
$A(e,e'p)$ scattering, which are missed in the conventional DWIA.

In our recent publication \cite{Helium4} we have given a simple
classification and evaluation
of the leading ISC and FSI contributions to the
missing momentum distribution in $^4He(e,e'p)$.
We also pointed out the numerically
very substantial novel effects of quantum-mechanical interference
between ISC and FSI terms and of the slowly decreasing FSI-induced
tail of the missing momentum distribution in longitudinal
kinematics, both of which defy the semiclassical description
(for a detailed discussion of the latter effect in
$A(e,e'p)$ scattering on heavier nuclei see \cite{NSZ}).
The finding \cite{Helium4} of dominant pure FSI and large
FSI-ISC interference
effects in the leading contributions to the missing momentum
distribution make a complete analysis of FSI and ISC
effects for $^{4}He(e,e'p)$ scattering a pressing issue.
In this communication, we report the results from such an
exploratory study of the
missing momentum distribution over the whole range of the
missing momentum $\vec{p}_{m}$, starting
with the realistic Jastrow correlated wave function and with the
Glauber
theory multiple scattering expansion for final state interaction
of the struck proton with the spectator nucleons.
Such an analysis is called upon
for several reasons. On the one hand, $^{4}He$ is a simple
enough nucleus in which the missing momentum distribution can be
calculated to all orders in the FSI and pair correlation function,
although such a calculation is quite a formidable task. Such a
calculation is indispensable for understanding to which extent
the FSI-modified missing momentum distribution, measured in $A(e,e'p)$
scattering, allows a reliable extraction of the SPMD and of the
short range correlation effects in the ground state of the target
nucleus. On the other hand, despite being a four-body system,
$^{4}He$ is a high density nucleus, and there are good reasons
to believe that an exhaustive analysis of FSI effects in
$^{4}He$  gives a good guidance to significance of FSI effects
in heavier nuclei. Finally, such an analysis allows to test
how much nuclear surrounding changes predictions from generally
accepted approximations, for instance, the dominance \cite{zabo} of
the
large-$p_{m}$ distributions by the quasi-deuteron configurations
in the nuclear medium. The case for principal FSI effects is solid
and the generality of our principal conclusions is not limited
by the model wave functions, the somewhat simplified
functional form of which was motivated by a pressing necessity
to circumvent the enormous complexity of numerical calculations.

The main conclusion of the present study is that distortions of
the missing momentum distribution by final state interactions are
very
strong over the whole phase space and make an
unambiguous, model independent, extraction of the large-$p_{m}$
component of the SPMD from the experimental data on $A(e,e'p)$
scattering hardly possible, perhaps impossible altogether.
The origin of this FSI dominance and parameters which control
this dominance, are well understood.
We also comment on the determination of nuclear transparency, on
the accuracy of the quasi-deuteron approximation, on the r\^ole of
FSI effects in a comparison
of the large-$p_{m}$ spectra for the deuteron and $^{4}He$
targets and on implications of FSI effects for
the $y$-scaling analysis. The analysis we report here
suggests unequivocally an even stronger dominance of FSI effects
in $A(e,e'p)$ scattering in heavier nuclei.




\section{Missing momentum distribution:
kinematics and definitions}

We wish to focus on FSI effects, and for the sake of simplicity we
consider the longitudinal response and treat the photon as a scalar
operator. Then, following the usual procedure of factoring out
the $ep$ scattering cross section \cite{frulmo,deForest}, the
experimentally
measured $A(e,e'p)$ coincidence cross section can be represented
in the form
\beq
{d\sigma \over dQ^{2}d\nu dp d\Omega_{p}}=
K|M_{ep}|^{2}S(E_{m},\vec{p}_{m},\vec p)\, .
\label{eq:2.1}
\endeq
Here $K$ is a kinematical factor, $M_{ep}$ is the $ep$ elastic
scattering amplitude, $\nu$ and $\vec{q}$ are the $(e,e')$ energy
and momentum transfer, $Q^{2}=\vec{q}\, ^{2}-\nu^{2}$, the struck
proton has a momentum $\vec{p}$ and energy $E(p)=T_{kin}+m_{p}$,
and the missing
momentum and the missing energy are defined as
$\vec{p}_{m}=\vec{q}-\vec{p}$ and $E_{m}=\nu+m_{p}-E(p)-T_{kin}(A-1)\,$
(where $T_{kin}(A-1)$ is the kinetic energy of the undetected $(A-1)$
residual system).
Hereafter, the $z$-axis is chosen along the $(e,e')$ momentum
transfer $\vec{q}$.
The spectral function $S(E_{m},\vec{p}_{m},\vec p)$ can be written in
terms of the nuclear reduced amplitudes ${\cal M}_f$ as (for instance,
see  \cite{Ciofi91})
\beq
S(E_{m},\vec{p}_{m})= \sum_{f} |{\cal M}_{f}|^{2}\delta(\nu+m_p-
E(p)-E_{m})\,.
\label{eq:2.2}
\endeq
For the longitudinal response, the reduced nuclear amplitude for
the exclusive process $^{4}He(e,e'p)A_{f}$ equals
\arr
{\cal M}_{f}=
\int d\vec{R}_{1}\,d\vec{R}_{2}\,d\vec{R}_{3}
\Psi_{f}^{*}(\vec{R}_{1},\vec{R}_{2})
\hat{S}(\vec{r}_{1},...,\vec{r}_{4})
\Psi(\vec{R}_{1},\vec{R}_{2},\vec{R}_{3})
\exp(i\vec{p}_{m}\vec{R}_{3})  \, .
\label{eq:2.3}
\endarr
Here $\Psi(\vec{R}_{1},\vec{R}_{2},\vec{R}_{3})$ and $
\Psi_{f}(\vec{R}_{1},\vec{R}_{2})$ are wave functions of
the target $^{4}He$ nucleus and of the specific 3-body final state
$A_{f}$, which are conveniently  described in terms of the
Jacobi coordinates $\vec{R}_{1}=\vec{r}_{2}-\vec{r}_{1}$,
{}~$\vec{R}_{2}={2\over 3}\vec{r}_{3}-{1\over 3}(\vec{r}_{1}+
\vec{r}_{2})$,~$\vec{R}_{3}=\vec{r}_{4}-{1\over 3}(\vec{r}_{1}+
\vec{r}_{2}+\vec{r}_{3})$ (plus $\vec R_{cm}  = {1\over 4}
\sum_i \vec r_i  \equiv  0$). Lab coordinates
$\{\vec r_i(\vec R_j,\vec R_{cm})\}$
are also used when appropriate.
The specific expression (\ref{eq:2.3}) for ${\cal M}_{f}$ corresponds
to the nucleon ``4'' of $^{4}He$ being
chosen for the detected struck proton.
$\hat{S}(\vec{r}_{1},...,\vec{r}_{4})$ stands for the
$S$-matrix of the
FSI of the struck proton with three spectator nucleons.

The calculation of the full FSI-modified spectral function
$S(E_{m},\vec{p}_{m})$ is a separate problem, which goes beyond
the scope of the present communication. In this exploratory study of
the salient features of FSI, we focus on the inclusive
missing momentum
spectrum of protons in $^{4}He(e,e'p )$ scattering in quasielastic
kinematics
\beq
W(\vec{p}_{m})={1 \over (2\pi)^{3}} \int  dE_{m} S(E_{m},\vec{p}_{m})
= {1 \over (2\pi)^{3}}\sum_{f} |{\cal M}_{f}|^{2}
\label{eq:2.4}
\endeq
Evidently, having the high kinetic energy of the struck proton
$T_{kin}$ is important for exhausting the missing energy spectrum
which at large $p_{m}$ is expected to extend to particularly
high $E_{m}$ [6]. This for the first time becomes possible
in the CEBAF range of $T_{kin}$.
The sum over all the allowed final
states $A_{f}$
for the three undetected nucleons can be performed making use of
the closure relation
\beq
\sum_{f}
\Psi_{f}(\vec{R}_{1}\,',\vec{R}_{2}\,')
\Psi_{f}^{*}(\vec{R}_{1},\vec{R}_{2}) =
\delta(\vec{R}_{1}-\vec{R}_{1}\,')
\delta(\vec{R}_{2}-\vec{R}_{2}\,')\,.
\label{eq:2.5}
\endeq
This leads to the missing momentum distribution
\beq
W(\vec{p}_{m}) = {1\over (2\pi)^{3}}\int
 d\vec{R}_{3}\,'d\vec{R}_{3}
 \rho(\vec{R}_{3},\vec{R}_{3}\,')
\exp\left[i\vec{p}_{m}(\vec{R}_{3}-\vec{R}_{3}\,')\right]
\, ,
\label{eq:2.6}
\endeq
where
\beq
 \rho(\vec{R}_{3},\vec{R}_{3}\,')
=
\int d\vec{R}_{1}\,d\vec{R}_{2}
\,  \Psi^{*}(\vec{R}_{1},\vec{R}_{2},\vec{R}_{3}\,')
S^{\dagger}(\vec{r}_{1}\,'..,\vec{r}_{4}\,')
S(\vec{r}_{1},...,\vec{r}_{4})
\Psi(\vec{R}_{1},\vec{R}_{2},\vec{R}_{3})\,
\label{eq:2.7}
\endeq
is the FSI-modified one-body density matrix (OBDM). In the PWIA, when
$\hat{S}(\vec{r}_{1},...,\vec{r}_{4})=1$, eq.~(\ref{eq:2.7})
reduces to the standard one-body density matrix of a nucleus
and eq.~(\ref{eq:2.6})
gives the familiar
SPMD  $N_{F}(p_{m})$, also often referred to
as the PWIA missing momentum distribution.
Extensive studies of $N_{F}(p_{m})$
are available in the literature
(\cite{Ciofi91,Traini}
and references therein, see also the monograph \cite{Antonov}).

The main goal of this paper is to investigate the influence of FSI on,
and the interplay of FSI and ISC effects in,
the missing momentum distribution $W(\vec p_m)$, in particular at
large missing momenta.
The calculation of a realistic $^4He$ wave function is a field of its
own. An accurate incorporation of FSI effects into the calculation of
the momentum distribution $W(\vec{p}_{m})$, Eq.~(\ref{eq:2.6}),
is a numerically very
involved problem. For this reason, in this exploratory study of FSI
effects,
we confine ourselves to a simple, yet realistic, parameterization of
the wave function, which allows a semi-analytic
evaluation of $W(\vec{p}_{m})$. Namely, we take the popular Ansatz
consisting of a harmonic oscillator mean field wave function $\Psi_o$
and Jastrow correlation factor $\hat{F}$ to allow for the ISC effects,
\beq
\Psi(\vec{R}_{1},\vec{R}_{2},\vec{R}_{3})\equiv\ \hat{F}\,
\Psi_{o}(\vec{R}_{1},\vec{R}_{2},\vec{R}_{3}) \,,
\label{eq:2.8}
\endeq
where
\beq
\hat{F} \equiv\
\prod_{i<j}^{4}\Big[1-C(\vec{r}_{i}-\vec{r}_{j})\Big]\, ,
\label{eq:2.9}
\endeq
\begin{equation}
\Psi_{o} \propto
\exp \Big[-{1\over 2R_{o}^{2}}\sum_{i}^{4} \vec{r}_{i}\,^{2} \Big]
=
\exp \Big[-{1\over 4R_{o}^{2}}\left(\vec{R}_{1}\,^{2}+
3\vec{R}_{2}\,^{2}+{3\over 2}\vec{R}_{3}\,^{2}\right)
\Big]
\label{eq:2.10}
\endeq
and
\begin{equation}
C(r)=C_{o}\exp\left(-{r^{2}\over 2r_{c}^{2}}\right).
\label{eq:2.11}
\end{equation}
For a hard core repulsion $C_{o}=1$, for a soft core $C_{o} < 1$.
In the literature, one usually considers the correlation
radius $r_{c}\approx 0.5$-$0.6$\,fm \cite{Traini,Antonov,Kezer,
Pieper,Co}. The radius $R_{0}$ of the
oscillator wave function can be determined from the charge radius
of the $^{4}He$, from which one must subtract the contribution
from the finite charge radius of the proton. For instance, for
a hard core repulsion, $C_{o} = 1$, and $r_{c}=0.5$\,fm,
the experimentally measured charge radius of the $^4He$
\cite{devries}
is reproduced with $R_{o} = 1.29$\,fm. Notice,
that the "correlation
volume" $r_c^3$ is a very small fraction of the volume of a nucleus
and
we have the
strong inequality $(r_{c}/R_{0})^{3} \ll 1$.




\section{Short range correlations and single particle momentum
distribution}

In order to set up a background, we start with a discussion
of the short range correlation effects in the SPMD
$N(\vec{p}_{m})$.
Here one
has to evaluate the conventional OBDM
\beq
\rho(\vec{R}_{3},\vec{R}_{3}\,')
=
\int d\vec{R}_{1}\,d\vec{R}_{2}
\, \Psi_{o}^{*}(\vec{R}_{1},\vec{R}_{2},\vec{R}_{3}\,')
\hat{F}^{\dagger}(\vec{r}_{1}\,'..,\vec{r}_{4}\,')
\hat{F}(\vec{r}_{1},...,\vec{r}_{4})
\Psi_{o}(\vec{R}_{1},\vec{R}_{2},\vec{R}_{3})\,
\label{eq:3.1}
\endeq
The product of the Jastrow functions in (\ref{eq:3.1}) can be
expanded as
\arr
\hat{F}^{\dagger} \hat{F}=
\prod_{i<j}^{4}\Big[1-C^{\dagger}(\vec{r}_{i}\,'-\vec{r}_{j}\,')\Big]
\Big[1-C(\vec{r}_{i}-\vec{r}_{j})\Big] =\nonumber \\
1-\sum_{i<j}\left[C^{\dagger}(i'-j')+C(i-j)\right] +
\sum C^{\dagger}C
+ \sum C^{\dagger}C^{\dagger}
+ \sum CC+ \cdots
\label{eq:3.2}
\endarr
Altogether there are $2^{12}$ terms in the integrand of the OBDM
in Eq.~(\ref{eq:3.1}), but with the considered Ansatz wave function
it is possible to carry out all the
integrations in the OBDM and in the SPMD
analytically. Here we present a brief summary of main effects of
short range correlations in the SPMD
(for the related earlier works see
\cite{Traini,Antonov,Kezer}, the more refined form of
the above approximation known as the
correlated basis function theory
is widely being applied to SPMD in heavier nuclei
(\cite{Co} and references therein)).

To the zeroth order in the correlation function, one finds the
long ranged
\footnote{
Hereafter, ''long ranged'' refers to functions which change
on the scale of the $^4He$ radius $R_{0}$, while ``short
ranged'' refers to functions which
change on the scale of the short range
correlation radius $r_c$ and the radius of the final state
proton-nucleon interaction $b_{0}$.}
 OBDM
\beq
\rho_{0}(\vec{R},\vec{R}') =w_{1} \, \frac{27}{512} \,  \frac{1}{\pi^3}
 \, {1\over R_{0}^{6}} \,
\exp \left ( -\frac{3}{8R_{o}^{2}}\left ( \vec{R}\, ^2+\vec{R'}\, ^2
\right )
\right )
\label{eq:3.3}
\endeq
(where we have introduced $\vec R \equiv \vec R_3$ and $\vec R' \equiv
\vec R_3'$ for brevity)
and
the steeply decreasing SPMD
\begin{equation}
N(1;\vec{p}_{m})= w_{1}\exp\Big(-{4\over 3}R_{o}^{2}p_m^2\Big),
\label{eq:3.4}
\end{equation}
falling off rapidly for momenta larger
than the Fermi momentum $k_{F} \sim 1/R_{0}$.
(Hereafter $N(1;\vec{p}_{m})$, $N(C;\vec{p}_{m})$,... indicate
the contributions to
$N(\vec{p}_{m})$ coming from the ``1'',``C'',... terms
in the expansion (\ref{eq:3.2})).
Note that in any PWIA calculation, with or without correlations,
the SPMD is isotropic and depends only on $|\vec p_m|$.

The higher order terms in $C,C^{\dagger}$ in (\ref{eq:3.2})
can be considered as "interactions" which modify the OBDM
$\rho_{0}(\vec{R},\vec{R}')$ of the mean field
approximation. This is shown schematically in Fig.~\ref{fig1}.
The first order corrections to $\rho(\vec{R},\vec{R}')$ and
$N(\vec{p}_{m})$ come from the terms which are
linear in $C(\vec{r}_{i}-\vec{r}_{j})$ and
$C^{\dagger}(\vec{r}_{i} \, ^{'}-\vec{r}_{j} \, ^{'})$. Leading
corrections to $N(\vec{p}_{m})$, which decrease with $p_{m}$
substantially
slower than the zeroth order term (\ref{eq:3.4}), come from
interactions of Fig.~\ref{fig1}a, which affect only one of the
trajectories in the calculation of the OBDM.
One can easily
verify that the corresponding corrections to the OBDM
are long-ranged, and lead to the contribution to
SPMD of the form
\begin{equation}
N(C^{\dagger}+C;\vec{p}_{m})\approx -6w_{1} C_{0}
\sqrt{\frac {27} {125}}
\Big({r_c \over R_o}\Big)^3
\exp\left (-{4 \over 5}R_{o}^{2}p_{m}^{2}\right ).
\label{eq:3.5}
\end{equation}
For the sake of brevity, we don't show here
and in the following equations correction
factors $[1+{\cal O}(r_{c}^{2}/R_{o}^{2}) ]$ to the slope and the
normalization factors. Of course, those factors are
included in our calculations. Notice, that the
normalization in (\ref{eq:3.5}) contains the small factor
$(r_{c}/R_{0})^{3}$. Notice also the destructive interference
between the $N(1;\vec{p}_{m})$ and $N(C^{\dagger}+C;\vec{p}_{m})$,
which becomes stronger with increasing $p_{m}$, because the latter
has a smaller slope of the $\vec{p}_{m}\,^{2}$ dependence than the
former.

The driving contributions to the short ranged component of the OBDM
and the related large-$p_{m}$ component of the SPMD come from the
three identical terms of the form
$C^{\dagger}(\vec{r}_{4}\,'-\vec{r}_{i}\,')
C(\vec{r}_{4}-\vec{r}_{i})$.  They produce a short range interaction
of the type shown in Fig.~\ref{fig1}b between the two trajectories
which enter the calculation of the OBDM and lead to a short ranged
component of the OBDM of the form
\arr
\rho(C^{\dagger}C,\vec{R},\vec{R}') = 3\int d^3 \vec R_1 d^3 \vec R_2
\, \Psi_o \Psi_o' \, C^{\dagger}(\vec{r_3}'-\vec{r_4}') \,
C(\vec{r_3}-\vec{r_4}) \nonumber\\ \approx w_{1} \, C_o^2 \,
\frac{1}{\pi^{3}}\, \sqrt{ \frac{3^{11}}{2^{21}}} \, \frac
{r_c^3}{R_{o}^{9}} \exp \left( - \frac{1}{4 r_c^2} \left( \vec R -
\vec R' \right)^2 - \frac{9}{8 R_o^2} \left( \vec R^2 + \vec R'^2
\right) \right) \, .
\label{eq:3.6}
\endarr
The resulting large-$p_{m}$ component of the SPMD directly
probes the correlation function $C(\vec{R})$:
\arr
N(C^{\dagger}C;\vec{p}_{m}) \approx
w_{1} C_o^2
\frac{1}{2 \pi ^3} \, \frac{1}{R_o^6} \,{\sqrt{\frac{243}{512}}       }
\left|\int d^{3}\vec{r}
C(\vec{r}) \exp(i\vec{p}_{m}\vec{r})\right|^{2}\nonumber\\
 =
N(C^{\dagger}C;\vec{p}_{m}) \approx
w_{1} C_{o}^{2} \sqrt {\frac{243}{512}}
\left ({r_c \over R_o} \right)^{6}
\exp\left(-r_{c}^{2}
p_{m}^{2}\right)\, .
\label{eq:3.7}
\endarr
(For the sake of brevity, we suppressed here the correction
factors $[1+{\cal O}(r_{c}/R_{0})^{2}]$ to the slope of
$p_{\perp}^{2}$ dependence
and
the total normalization.)
Notice, that the short ranged component of the OBDM
is proportional to the correlation volume, whereas the
large-$p_{m}$ tail of the SPMD contains
the small normalization factor $\propto
(r_{c}/R_{o})^{6}$.

The correlated pair of nucleons is often treated as a
quasi-deuteron (\cite{Ciofi91} and references therein).
Indeed, Eq.~(\ref{eq:3.7}) resembles the
momentum distribution in the quasi-deuteron with the
short-range correlation function playing the r\^ole of
the wave function of the quasideuteron (for the recent
analysis of $^{2}H(e,e'p)$ scattering see
\cite{Deuter,Tensor}).

The terms $\propto CC, C^{\dagger}C^{\dagger}$ in expansion
(\ref{eq:3.2}) correspond to interactions which involve only
one of the trajectories in $\rho(\vec{R},\vec{R}')$ and give a
long ranged contribution to the OBDM and a
steeply decreasing contribution to the SPMD $N(\vec{p}_{m})$,
similar to (\ref{eq:3.4}),(\ref{eq:3.5}).
The above analysis can easily be extended to still higher order
effects in the correlation function. For instance, the
cyclic terms of the form $C^{\dagger}(\vec{r}_{4}-\vec{r}_{3})
C(\vec{r}_{4}^{'}-\vec{r}_{2})C(\vec{r}_{2}-\vec{r}_{3})$
also lead to a short-range interaction between the two
trajectories in the OBDM, as shown in Fig.~\ref{fig1}c.
For the presence of the extra link in the 4-2-3 chain,
though, the corresponding interaction range
is larger than in the diagram of Fig.~\ref{fig1}b, and the resulting
contribution from such cyclic terms
to the SPMD has a $p_{m}$ dependence steeper than
in Eq.~(\ref{eq:3.7}):
\begin{equation}
N(\left[ C^{\dagger}CC+C^{\dagger}C^{\dagger}C\right]_{cyclic};
\vec{p}_{m}) \approx
- 6w_1 C_o^3 \sqrt{ \frac{1}{8}} \left( \frac {r_c}{R_o} \right)^9
\exp(-{3\over 2}r_{c}^{2}p_{m}^{2})
\label{eq:3.8}
\end{equation}
However,
numerically more important are the non-cyclic third order
terms of the form
$C^{\dagger}(\vec{r}_{4}^{'}-\vec{r}_{i}^{'})
C(\vec{r}_{4}-\vec{r}_{i})
C^{\dagger}(\vec{r}_{j}-\vec{r}_{k})$, which have a slow
$p_{m}^{2}$ dependence equal to that of the leading $C^{\dagger}C$
contribution (\ref{eq:3.6}) but are of the opposite sign:
\begin{equation}
N(\left[ C^{\dagger}CC+C^{\dagger}C^{\dagger}C\right]_{non-cyclic};
\vec{p}_{m}) \approx -10
w_{1} C_{o}^{3} \sqrt {\frac{243}{512}}
\left ({r_c \over R_o} \right)^{9}
\exp\left(-r_{c}^{2}
p_{m}^{2}\right)\, .
\end{equation}
Such a destructive interference between the
$C^{\dagger}CC+C^{\dagger}C^{\dagger}C$ and
$C^{\dagger}C$ contributions to the
SPMD
demonstrates a well understood suppression
of interaction between any pair of nucleons for the repulsive
correlation with surrounding nucleons, for more
discussion see below.

The above considerations are illustrated by the numerical results
shown in Figs.\ref{figpwc},\ref{figpwrc},\ref{figpwcon},\ref{figrc}.
The
sensitivity of the SPMD to short range correlations is obvious from a
comparison in Fig.\ref{figpwc} of the mean field distribution,
$C_{o}=0$, with the soft core correlation $C_{o}=0.5$ and the hard
core correlation $C_{o}=1$.  For $C_{o} = 0$, the mean field SPMD
(\ref{eq:3.4}) falls off rapidly for $p_{m} \gsim k_{F} \sim 1/R_{o}$.
Once the short range correlations are included, the large-$p_{m}$ tail
builds up, with the strength which closely follows the law $\propto
C_{o}^{2}$. The effect of destructive interference between the mean
field SPMD (\ref{eq:3.4}) and the lowest order ISC contribution
(\ref{eq:3.5}) is obvious at intermediate $p_{m}\sim k_{F}$. The
interference effect rises with the correlation strength $C_{o}$. One
usually considers $r_{c} \sim 0.5$\,fm, but the exact value of the
correlation radius $r_{c}$ is not known precisely
\cite{Traini,Kezer,Pieper}.  According to the estimate (\ref{eq:3.7}),
the ISC contribution to SPMD rises steeply, $\propto r_{c}^{6}$, with
the correlation radius $r_{c}$. However, at large values of $p_{m}$
where the short range correlation component takes over the mean field
component, the suppression by the factor $\exp(-r_{c}p_{m}^{2})$ is
quite significant and the residual sensitivity to the variation of
$r_{c}$ is weaker than $\propto r_{c}^{6}$. This is illustrated by
Fig.\ref{figpwrc}, where we compare the SPMDs for $r_{c}=0.5$\, fm and
$r_{c}=0.6$\,fm.

Above we cited the explicit form of leading terms in the SPMD.  The
rate of convergence of the expansion in powers of the correlation
function is of great interest on its own. For instance, the
large-$p_{m}$ tail of the SPMD is often discussed in terms of the
quasi-deuteron configurations, neglecting the effect of the nuclear
surrounding. If this were a good approximation, then one would have
expected a universality of the large-$p_{m}$ SPMDs for all nuclei.
The accuracy of this approximation can be judged from
Fig.\ref{figpwcon}, where we present the results for $N(p_m)$
including all terms up to the second order, i.e.,
$C^{\dagger}C^{\dagger}+C^{\dagger}C+CC$ (dotted line), to the fourth
order (dashed line), to the sixth order (dash-dotted line), to the
eighth order (double-dotted line), and the full expansion
(\ref{eq:3.2}) up to the twelfth order terms (solid line) is
plotted. (Truncation of an expansion at lowest
odd orders of the correlation
function can result in $N(\vec{p}_{m})$ which is not positive
valued). All the different single-particle momentum distributions are
normalized to unity.  Although the shape of the SPMD does not change
qualitatively, we find a substantial, $\sim 40\%$,
renormalization of the lowest-order quasi-deuteron
contribution to $N(\vec{p}_{m})$ by effects of the nuclear
environment.
This shows that the large-$p_{m}$ behaviour of the SPMD
is sensitive to details of the nuclear wave function and a
quasi-deuteron universality of the SPMD
at large $p_{m}$ is though a reasonable, but not a very
accurate, approximation.
The effect of higher order corrections is evident from
the form of the Jastrow correlation factor: the mutual repulsion from
surrounding nucleons suppresses the probability of short distance
interaction between any pair of nucleons in the nucleus, cf.
Eqs.~(\ref{eq:4.8}) and (\ref{eq:4.9},\ref{eq:4.10}).
The effect of higher order terms is particularly
important at intermediate values of $p_{m}\sim k_{F}\sim 1/R_{0}$,
where our model SPMD develops a slight minimum.
Our model wave function
contains only those many nucleon correlation effects, which are
reducible to higher orders in the pair correlation function. The fact
that terms up to sixth order are non-negligible
at large $p_{m}$, suggests that the
presently poorly known pair-irreducible multinucleon correlations can
also contribute to the SPMD at large momentum.

Fig.~\ref{figpwc} shows that the shape of the SPMD changes with the
correlation strength in a nontrivial way.
In Fig.\ref{figrc}  we present the renormalization factor
\begin{equation}
R_{C}(p_{m})=
{N(C_{0}=1;p_{m})\over N(C_{0}=0;p_{m})}\, ,
\label{defrc}
\end{equation}
which expands to greater detail the difference  between the
mean field and correlated distributions.
At large $p_{m}$ this ratio blows up,
but at intermediate $p_{m} \lsim k_{F}$ it has a nontrivial
behaviour:
The short range
correlation effects {\sl enhance, rather than suppress},
the SPMD and lead to $R_{C}(p_{m}) > 1$ at small $p_{m}$,
the extra strength at large $p_{m}$ due to short range
correlations, comes from a depletion of SPMD at $p_{m}\sim k_{F}$,
rather than from the region of $p_{m}\sim 0$.

There are no direct experimental data on the SPMD in $^{4}He$ to
compare
our results with. Our model SPMD
$N(\vec{p}_{m})$ is close to the results from the recent Monte Carlo
calculation \cite{wiringa92} and the $y$-scaling analysis
\cite{Ciofi91}.




\section{Final state interaction effects.}

At the large $Q^{2}$ of interest in the CEBAF experiments, FSI
can be described by Glauber theory.
Defining transverse and longitudinal components
$\vec{r}_{i}$ $\equiv$
$(\vec{b}_{i},z_{i})$ and $\vec{R}_{i}$ $\equiv$
$(\vec{B}_{i},Z_{i})$ we can write
\beq
\hat{S}(\vec{r}_{1},...,\vec{r}_{4}) = \prod_{i=1}^{3}
\Big[ 1-\theta(z_{i}-z_{4})\Gamma(\vec{b}_{4}-\vec{b}_{i})\Big],
\label{eq:4.1}
\endeq
where $\Gamma(\vec{b})$ is the profile function of the
nucleon-nucleon scattering
\beq
\Gamma(\vec{b})
\ \equiv\
{ \sigma_{tot} (1 - i \rho) \over 4 \pi b_{o}^2  }
\exp \Big[-{\vec{b}^2 \over 2 b_{o}^2} \Big]
\label{eq:4.2}
\endeq ($\rho$ is the ratio of the real to imaginary part of the
forward elastic scattering amplitude). The Glauber formalism describes
quite well nucleon-nucleus scattering at energies above 500 MeV,
even at angles as large as 30$^o$ at 500 MeV (for a review see
\cite{abv78}). At $T_{kin} \sim 1$GeV, the experimental data on
$pN$ scattering give
$b_{o}\approx 0.5 fm$,
$\sigma_{tot} \approx 40mb$  and $\rho = 0.33$
\cite{abv78,lasi,lll93}. These parameters of $pN$ scattering
vary only very weakly over the GeV energy range relevant
to the CEBAF experiments. This weak dependence of
FSI on the kinetic energy of the proton has actually been
used in the application of closure to derivation of Eq.~(\ref{eq:2.7})
for the FSI-modified OBDM, in which we neglected the dependence
of the FSI operator $S(\vec{r}_{1},...,\vec{r}_{4})$ on the
missing energy $E_{m}$.

Combining together the FSI and the Jastrow correlation factors, we
can write down the FSI-modified density matrix as
\arr
\rho(\vec{R}_{3},\vec{R}_{3}\,')
=
\int d\vec{R}_{1}\,d\vec{R}_{2} ~~~~~~~~~~~~~~~~~~~~~~~~~~~~~~~~~~
\nonumber\\
\times \Psi_{o}^{*}(\vec{R}_{1},\vec{R}_{2},\vec{R}_{3}\,')
\hat{F}^{\dagger}(\vec{r}_{1}\,'..,\vec{r}_{4}\,')
\hat{S}^{\dagger}(\vec{r}_{1}\,'..,\vec{r}_{4}\,')
\hat{S}(\vec{r}_{1},...,\vec{r}_{4})
\hat{F}(\vec{r}_{1},...,\vec{r}_{4})
\Psi_{o}(\vec{R}_{1},\vec{R}_{2},\vec{R}_{3})\,.
\label{eq:4.3}
\endarr
The operator $\hat{F}^{\dagger}\hat{S}^{\dagger}\hat{S}\hat{F}$
which emerges in (\ref{eq:4.3}) can be expanded as
\begin{eqnarray}
\hat{F}^{\dagger}\hat{S}^{\dagger}\hat{S}\hat{F}
{}~ = ~
\prod_{i<j}^{4}\Big[1-C^{\dagger}(\vec{r}_{i}\,'-\vec{r}_{j}\,')\Big]
\Big[1-C(\vec{r}_{i}-\vec{r}_{j})\Big] \nonumber \\
\times
\prod_{i\neq 4}
\Big[ 1-\theta(z_{i}'-z_{4}')\Gamma^{\dagger}
(\vec{b}_{4}\,'-\vec{b}_{i}\,')\Big]
\Big[ 1-\theta(z_{i}-z_{4})\Gamma(\vec{b}_{4}-\vec{b}_{i})\Big]=
\nonumber\\
1-\sum_{i<j}\left[C^{\dagger}+C\right]
-\sum_{i\neq 4}
\left[\Gamma^{\dagger} +
\Gamma \right]
+ \sum [C^{\dagger}\Gamma +C\Gamma^{\dagger}]
+ \sum C^{\dagger}C + \sum \Gamma^{\dagger} \Gamma +. ...
\label{eq:4.4}
\end{eqnarray}
The r\^ole of the FSI terms in the expansion (\ref{eq:4.4}) is
very similar to that of the short-range correlation terms.
There are interactions which only involve one of the trajectories
in the FSI-modified OBDM, there are
terms of the form
$\theta(z_{i}'-z_{4}')
\theta(z_{i}-z_{4})
\Gamma^{\dagger}(\vec{b}_{4}^{'}-\vec{b}_{i}^{'})
\Gamma(\vec{b}_{4}-\vec{b}_{i})$, which lead to an
interaction between the two trajectories etc. There also emerges
a novel kind of an interaction between the two trajectories
of the form
$C^{\dagger}(\vec{b}_{4}^{'}-\vec{b}_{i})
\Gamma(\vec{b}_{4}-\vec{b}_{i})$ (plus its hermitian conjugate
and higher order short ranged
interactions), which is due to the quantum-mechanical
interference between the initial state correlations and final state
interactions \cite{Helium4}.
The r\^ole of different terms in the expansion (\ref{eq:4.4}) will
be discussed in more detail below.

The salient features of FSI stem from the observation that the
Glauber operator $\theta(z_{i}-z_{4})\Gamma(\vec{b}_{4}-\vec{b}_{i})$
is a short ranged function of the transverse separation
$\vec{b}_{4}-\vec{b}_{i}$ and a long ranged function of the
longitudinal separation $\theta(z_{i}-z_{4})$.
This angular anisotropy of the FSI operator leads to an
angular anisotropy of the missing momentum distribution $W(\vec
p_m)$ and in the further discussion we
decompose the missing momentum
into transverse and longitudinal components
$\vec p_m = (\vec p_{\perp},p_{m,z})$.

 In the range of kinetic energies of the proton typical of
the CEBAF range of $Q^{2}$,
the transverse range of FSI, $b_{0}\approx 0.5$\,fm,
is numerically very close to the correlation
radius: $b_{0}\sim r_{c}$. Now we briefly recapitulate the
most striking effects of FSI terms in the expansion (\ref{eq:4.4}),
following the classification developed in \cite{Helium4}.

The terms linear in $\Gamma^{\dagger}$ and $\Gamma$ correspond to
an interaction with one of the two trajectories in the OBDM
as shown in Fig.~\ref{figdiagfsi}a.
As such, they lead to a long ranged contribution
to the OBDM and
to a contribution $W(\Gamma^{\dagger}+\Gamma; \vec{p}_{m})$
to the FSI-modified missing momentum distribution of the form
very similar to $N(C^{\dagger}+C;\vec{p}_{m})$ of Eq.~(\ref{eq:3.3}),
but with the normalization typical of the Glauber multiple
scattering theory \cite{Glauber,Dakhno}
\beq
Y\sim
{\sigma_{tot} \over 4\pi b_{0}^{2}}\cdot
\left({b_{0}\over R_{0}}\right)^{2}=
{\sigma_{tot} \over 4\pi R_{0}^{2}} \, ,
\label{eq:4.5}
\endeq
which is larger than the normalization in (\ref{eq:3.5})
by a factor of the form
\beq
{W(\Gamma^{\dagger}+\Gamma; \vec{p}_{m})\over
N(C^{\dagger}+C;\vec{p}_{m})} \sim
\left({\sigma_{tot} \over 4\pi b_{0}^{2}}\right)\cdot
\left({b_{0} \over R_{0}}\right)^{2} \cdot
\left({R_{0}\over r_{c}}\right)^{3} \cdot {1 \over C_{0}}
 \sim {R_{0} \over r_{c}}\cdot {1\over C_{0}} \gg 1
 \, .
\label{eq:4.6}
\endeq
Here we made use of the fact that $\sigma_{tot} \sim 4\pi b_{0}^{2}$
and $b_{0}\sim r_{c}$. The enhancement factor $\sim R_{0}/r_{c}$
in (\ref{eq:4.6}) derives from the fact that the Glauber operator
is a long ranged function of the longitudinal separation in
contrast to the short ranged correlation function and is the
simplest demonstration that FSI effects in the missing momentum
distribution are more important than the ISC effects.
Eq.~(\ref{eq:4.6}) gives an estimate of the relative normalizations
of the two components of the missing momentum distribution. The
transverse momentum dependence of both components is essentially
identical. For the presence of the $\theta$-function in the Glauber
operator, the $p_{m,z}$ dependence of $W(\Gamma;\vec{p}_{m})$ can not
be written down in a simple analytical form (one can derive an
involved and not very enlightening expression in terms of an
error function of a complex argument); what is important is that
it is a steeply decreasing function of $p_{m,z}$ on the scale $k_{F}$.

An analysis of higher order effects proceeds very similarly.
The driving short ranged FSI contribution to the OBDM
and to the large-$p_{m}$ tail of the missing
momentum distribution $W(\vec{p}_{m})$ comes from diagrams
of the type shown in Fig.~\ref{figdiagfsi}b which correspond to terms
$\propto \Gamma^{\dagger}
(\vec{b}_{4}^{'}-\vec{b}_{i}^{'})\Gamma(\vec{b}_{4}-\vec{b}_{i})$
in the expansion (\ref{eq:4.4}). Suppressing for a while the
$p_{m,z}$ dependence and making use of the strong inequality
$(b_{0}/R_{0})^{2} \ll 1$, in close similarity to
$N(C^{\dagger}C;\vec{p}_{m})$ we find
\arr
W(\Gamma^{\dagger}\Gamma;\vec{p}_{m}) \propto
\left|\int d^{2}\vec{B}
\Gamma(\vec{B})
\exp(i\vec{p}_{\perp}\vec{B})\right|^{2}
=
4\pi {d\sigma_{el} \over dp_{\perp}^{2}} =
{1 \over 4}\sigma_{tot}^{2}(1+\rho^{2})\exp(-b_{o}^{2}p_{\perp}^{2})
\, .
\label{eq:4.7}
\endarr
The result that this particular
contribution to the transverse missing
momentum distribution is proportional to the differential
cross section of elastic $pN$
scattering $d\sigma_{el}/dp_{\perp}^{2}$, is self-explanatory
and the effect of $\Gamma^{\dagger}\Gamma$ interaction admits
a quasiclassical interpretation as an incoherent elastic
rescattering of the struck proton on spectator nucleons.
The still higher order terms of the form
$\propto \Gamma^{\dagger}
(\vec{b}_{4}^{'}-\vec{b}_{i}^{'})\Gamma(\vec{b}_{4}-\vec{b}_{i})
 \Gamma^{\dagger}
(\vec{b}_{4}^{'}-\vec{b}_{k}^{'})\Gamma(\vec{b}_{4}-\vec{b}_{k})$
lead to a second order short ranged interaction between the
trajectories in the FSI-modified OBDM. The transverse momentum
dependence of such a contribution to
$W(\vec{p}_{m})$ also admits a semiclassical
form of the convolution of differential
cross sections of consecutive incoherent elastic scatterings and
\arr
W(\Gamma^{\dagger}\Gamma\Gamma^{\dagger}\Gamma;\vec{p}_{m}) \propto
Y^{2}\exp(-{1\over 2}b_{o}^{2}p_{\perp}^{2})
\, .
\label{eq:4.8}
\endarr The $\propto (b_{0}/R_{0})^{2}$ corrections to the slopes in
(\ref{eq:4.7}) can readily be derived and correspond very
qualitatively to a slight smearing of the elastic scattering cross
section due to the -compared to the momentum transfer in elastic $pN$
scattering- small initial Fermi motion of the struck proton and the
spectator nucleon (for a critical discussion of semiclassical
considerations see \cite{NSZ}).

Because of $b_{o}\approx r_{c}$ in  the CEBAF range of $Q^{2}$,
the short ranged $C^{\dagger}C$ and $\Gamma^{\dagger}\Gamma$
interactions between the two trajectories in $\rho(\vec{R},\vec{R}')$
lead to $W(C^{\dagger}C;\vec{p}_{m})=
N(C^{\dagger}C;\vec{p}_{m})$ and
$W(\Gamma^{\dagger}\Gamma;\vec{p}_{m})$
components with very similar $p_{\perp}$
dependence. In close similarity to
Eq.~(\ref{eq:4.6}), the overall normalization is substantially
larger for the FSI term. A comparison of the two components at
$p_{m,z}=0$ gives
\beq
{W(\Gamma^{\dagger}\Gamma;\vec{p}_{m}) \over
W(C^{\dagger}C;\vec{p}_{m})}
\approx
{1 \over C_{o}^{2}\sqrt{6}}\cdot
\left[{\sigma_{tot} \over 4\pi r_{c}^{2}}\right]^2\cdot
{\left(R_{o} \over r_{c}\right)^{2}} \sim 7\,
\label{eq:4.9}
\endeq
and leads to the important
conclusion that the tail of the transverse missing momentum
distribution must be entirely dominated by FSI effects.

Because the FSI is long-ranged in $z_4-z_i$,
$W(\Gamma^{\dagger}\Gamma;\vec{p}_{m})$ decreases
steeply with $p_{m,z}$ on the scale $\sim k_{F}\sim {1/R_{0}}
$. On the one hand, this leads to a strong
angular anisotropy of the elastic rescattering effect, cf.
Eq.~(\ref{eq:4.8}).
On the other hand, one would have naively expected the
continuation of this steep decrease of the $\Gamma^{\dagger}\Gamma$
contribution and the
dominance of the $C^{\dagger}C$ correlation component (\ref{eq:3.7})
of the SPMD in the
longitudinal kinematics at large $|p_{m,z}|\gsim k_{F}$.
There are deep quantum mechanical reasons why this is not the
case.
The $p_{m,z}$ dependence of $W(\Gamma^{\dagger}\Gamma;\vec{p}_{m})$
has peculiarities of its own. The $\theta$-function, which is
present in the Glauber operator, has a slowly decreasing Fourier
transform at large $p_{m,z}$.
The contribution from the
$\Gamma^{\dagger}\Gamma$ interaction to the integrand of the
OBDM contains the product of the $\theta$-functions
of the form
$\theta(z_{i}'-z_{4}')\theta(z_{i}-z_{4})=
\theta(z_{i}-z_{max})$,
where
\beq
z_{max}={1\over 2}(z_{4}+z_{4}^{'})+{1\over 2}|z_{4}-z_{4}^{'}|.
\label{eq:4.10}
\endeq
The nonanalytic function $|z_{4}-z_{4}^{'}|$
in $z_{max}$ gives rise to a $p_{m,z}^{-2}$ tail of the
longitudinal missing momentum distribution after the calculation
of the Fourier transform in $z_{4}-z_{4}^{'}$ (for a detailed
discussion of the case of heavy nuclei see [11]). The factor
$\theta(z_{i}'-z_{4}')\theta(z_{i}-z_{4})$ implies that
the $\Gamma^{\dagger}\Gamma$ interaction is forbidden on the part
of either one or the other of the trajectories in the calculation of
the OBDM as shown schematically in Fig.~\ref{figdiagfsi}.
This ban on the
$\Gamma^{\dagger}\Gamma$ interaction is of purely
quantum-mechanical origin and the $\propto p_{m,z}^{-2}$ tail
of
the longitudinal missing momentum distribution defies any
classical interpretation, in contrast to the strong FSI
enhancement of the transverse missing momentum distribution,
which admits a semiclassical interpretation to a certain extent.
For the reasons explained in detail in [11], the $\Gamma^{\dagger}
\Gamma$ interaction and the resulting $\propto p_{m,z}^{-2}$
tail of the missing momentum distribution are missed in the
conventional DWIA.

Finally, still another nontrivial
effect, the quantum-mechanical interference of
ISC and FSI,  comes from the terms $\propto
C^{\dagger}(\vec{r}_{4}\,'-
\vec{r}_{i}\,')\Gamma(\vec{b}_{4}-\vec{b}_{i})$ and $\propto
C(\vec{r}_{4}-\vec{r}_{i})
\Gamma^{\dagger}(\vec{b}_{4}\,'-\vec{b}_{i}\,')$. These and
higher order cyclic and non-cyclic terms lead to a short range
interaction as shown in Fig~\ref{figdiagfsi}c in the transverse
separation
between the trajectories in the calculation of the
OBDM (\ref{eq:4.1}) and to a large-$p_{m}$
tail of the transverse missing momentum distribution of
the form
\arr
w(C\Gamma^{\dagger}+C^{\dagger}\Gamma;\vec{p}_{m})\propto
\int d^{2}\vec{R}C^{\dagger}(\vec{R})\exp(i\vec{p}_{\perp}\vec{R})
\cdot
{\rm Re} \int d^{2}\vec{B}\Gamma(\vec{B})\exp(-i\vec{p}_{\perp}\vec{B})
\nonumber\\
\propto
\exp\left[-{1\over 2}(r_{c}^{2}+b_{o}^{2})p_{\perp}^{2}\right]\, .
\label{eq:4.11}
\endarr
Considering that $b_{0}\sim r_{c}$, at large $p_{\perp}$ the effects
of the $C\Gamma^{\dagger}+C^{\dagger}\Gamma$ interaction are
as important as those of the $C^{\dagger}C$ and
$\Gamma^{\dagger}\Gamma$ interactions. Furthermore,
the remarkable feature of the FSI-ISC interference term is that
owing partly to numerical factors, it has
a large normalization. At $p_{m,z}=0$ we find
\beq
{W(C\Gamma^{\dagger}+C^{\dagger}\Gamma;\vec{p}_{m})
\over
W(\Gamma^{\dagger}\Gamma;\vec{p}_{m})}\approx 4\sqrt{{3\over 5}}
C_{o}\left(4\pi r_{c}^{2} \over \sigma_{tot}\right)
\cdot {r_{c}\over R_{o}} \sim 1\,.
\label{eq:4.12}
\endeq
This clearly shows that the FSI-ISC interference effect is
much more important than the pure ISC component of the
missing momentum distribution.
Notice that any semiclassical
consideration would completely miss this large ISC-FSI
interference effect. Furthermore, the numerical
significance of this interference effects shows that even in
transverse kinematics, the semiclassical treatment of FSI
misses important quantum mechanical effects and must be taken
with great caution.
The $p_{m,z}$ dependence of this contribution to $W(\vec{p})$ will
be controlled by the long range character in the longitudinal
separation of the Glauber operator.




\section{Final state interaction and missing momentum distribution:
the numerical results}

The full calculation of the FSI-modified OBDM is quite involved,
because the full expansion (\ref{eq:4.3},\ref{eq:4.4}) for the
integrand of the OBDM involves $2^{18}$ terms. The
$\theta$-function in the Glauber operator leads to the
slow convergence of the Fourier transform and requires very
tight control of numerical accuracy in the calculation
of the FSI-modified missing momentum distribution, especially
at large $|p_{m,z}|$. With the generic wave function, an accurate
calculation of the large-$p_{m}$ behaviour of $W(\vec{p}_{m})$
would have required enormous computing time. Our Ansatz wave
function, complemented by the usual Gaussian form of the profile
function, simplifies the task greatly, because all the transverse
coordinate integrations and the corresponding Fourier transform
in the transverse missing momentum can be performed analytically
to all orders in ISC and FSI.
Only the longitudinal coordinate integrations and the corresponding
Fourier transform must be performed numerically.
Now we present some of the results for the missing momentum
distribution $W(\vec{p}_{m})$. Unless specified otherwise,
all the numerical results are for the hard core correlation,
$C_{0}=1$.

As the PWIA missing momentum distribution including the
correlation functions is completely isotropic, any anisotropy in
$W(\vec p_m)$ is a clear signal of FSI.  In Fig.~\ref{figad},
 we show the angular
distribution of $W(\vec p_m)$ for different missing momenta for the
full calculation with FSI (solid line) and the PWIA result (dashed
line). Already at the rather low missing momentum of $p_m = 1.0
fm^{-1}$, there is a strong deviation from the isotropic PWIA
behaviour. With growing missing momentum, the dip around $90^o$
evolves through a very asymmetric stage into a pronounced peak
at $p_{m} \gsim 1.6\,fm^{-1}$, the
signal of complete FSI (plus ISC-FSI interference)
 dominance.  The evolution of the angular distribution is
especially fast around $p_{m}\sim 1.5$\,fm$^{-1}$.
 One of the striking effects in
Fig.~\ref{figad} is the forward-backward asymmetry $W(
p_{\perp},p_z) \not=
W(p_{\perp}, -p_z)$.
This forward-backward asymmetry has its origin in the
nonvanishing real parts of the $p-n$ and $p-p$ scattering amplitudes
$\rho \not= 0$, leading to $
\hat{S}^{\dagger}(\vec b_1^{'}, z_1^{'}, \dots \vec b_4^{'}, z_4^{'})
\hat{S}(\vec b_{1}, z_{1}, \dots \vec b_{4}, z_{4}) \not=
\hat{S}^{\dagger}(\vec b_{1}, z_{1}, \dots \vec b_{4}, z_{4})
\hat{S}(\vec b_1^{'},z_1^{'}, \dots \vec b_4^{'}, z_4^{'})$
in the integrand of (\ref{eq:2.7}).

The same features of $W(\vec{p}_{m})$ can be seen in the
missing momentum distributions displayed for longitudinal and
transverse kinematics in Fig.~\ref{figmdwn}.
For small missing momentum, the FSI
leads to a reduction of strength, at $p_m = 0 $ the PWIA
distribution is depleted by $ \sim 24 \%$, which is about twice
as large as the nuclear shadowing effect in the total $p \: ^{4}He$
cross section \cite{Dakhno}.
Two mechanisms contribute to this depletion which is mostly due
to the $\Gamma^{\dagger}+\Gamma$ terms in the
expansion (\ref{eq:4.4}): i)
 attenuation of the flux of struck protons
due to absorption by inelastic interaction with the spectator nucleons,
ii) deflection of the struck protons by elastic scattering on
the spectator nucleons
Such a semiclassical interpretation
must be taken with the grain of salt, though: because of the
quantum-mechanical interference effects, FSI leads to quite an
involved, strongly $\vec{p}_{m}$ dependent, pattern of depletion
and/or enhancement of the FSI modified missing momentum distribution
$W(\vec{p}_{m})$ as compared to the PWIA distribution
$N(\vec{p}_{m})$ (see also below, section 7).
With the increasing missing momentum, up to $p_m < 0.5
fm^{-1}$, the angular distribution is almost isotropic, from $p_m =
0.5 fm^{-1}$ on, the deviations from the isotropic PWIA behaviour grow
larger.  In transverse kinematics, from $p_m = 1.5 fm^{-1}$ on, the
full calculation including FSI (solid line) shows a large tail at high
missing momenta, which is one order of magnitude larger than the tail
of the PWIA distribution $N(p_m)$ for hard core correlations
(dotted line).

The peculiarities of the Fourier transform of the $\theta$-function
factors do not show up in transverse kinematics, at $p_{m,z}=0$.
Here the r\^ole of the Glauber operator $\Gamma(\vec{b})$  in
the evaluation of the FSI-modified OBDM is nearly identical to the
r\^ole
of the correlation function $C(\vec{r})$ in the modification of
the mean-field OBDM. The only difference is in a much
larger strength
of FSI effects, which is the reason why $W(\vec{p}_{m})$ develops
a minimum at $p_{m}\approx 1.4\,fm^{-1}$ as compared to the
minimum of the PWIA distribution at $p_{m}\approx 1.7\,fm^{-1}$.

To this end, notice that $p_{m}\sim 1.4\,fm^{-1}$ corresponds to
a region of $p_{m}$ in which the effect of short range correlations
in the SPMD is still marginal. Already this observation
suggests very strongly
that in transverse kinematics, $\theta=90^{o}$,
the missing momentum distribution
$W(\vec{p}_{m})$ can only weakly depend on the short-range
correlation effects in the nuclear wave function. This is indeed
the case. The results shown in Fig.~\ref{figmdcpa1}  demonstrate
that, in a
striking contrast to the strongly correlation dependent
PWIA distribution $N(\vec{p}_{m})$
of Fig.~\ref{figpwc}, the FSI-modified $W(\vec{p}_{m})$ is extremely
insensitive to the strength of short range correlation. Namely,
even at large $p_m \gsim 1.5 fm^{-1}$
the missing momentum distribution $W(\vec p_m)$ calculated
with $C_o = 1$ is only by $\sim 50\%$ larger than $W(\vec p_m)$
calculated with $C_o = 0$, in contrast to a difference of several
orders in magnitude in the PWIA distributions for $C_o = 0$ and
$C_o = 1$.
This enhancement is much stronger
than expected from the ISC contribution of PWIA, though.
In a classical consideration, such
an enhancement of $W(\vec{p}_{m})$ with the correlation strength
is quite counterintuitive: the classical
probability of elastic rescattering
of the struck proton on the spectator nucleon is higher for close
configurations, which are suppressed by the hard core correlation.
The found enhancement of $W(\vec{p}_{m})$ from the mean field,
$C_{0}=0$, to the hard core correlation, $C_{0}=1$, wave function
must be attributed to the above discussed ISC-FSI interference
effect, see Eq.~(\ref{eq:4.10}), which is numerically larger than
the effect of suppression of $\Gamma^{\dagger}\Gamma$ elastic
rescattering contribution by the short range correlation
\cite{Helium4}.
Fig.~\ref{figmdcpa2}  shows that
the situation changes neither qualitatively nor
quantitatively if the correlation radius is
increased from $r_{c}=0.5\, fm$ to $r_{c}=0.6\,fm$.
Thus, we are led to the conclusion that
FSI is dominating completely
and that it would be extremely difficult to
disentangle the effects of short range correlations in the
ground state of the target nucleus
from the FSI-affected
experimental data on $A(e,e'p)$ scattering in
transverse kinematics.

Let's have a look at longitudinal kinematics, $\theta = 0^o$ and
$\theta = 180 ^o$, to check if the situation there is more promising.
In the angular distributions of Fig.~\ref{figad} one can see
that also in
longitudinal kinematics, deviations from the PWIA behaviour are
present and large. The missing momentum distributions
for longitudinal
kinematics displayed in Figs.\ref{figmdwn}b,c differ strongly
from $W(\vec p_m)$ in
transverse kinematics. Firstly, in the antiparallel kinematics, at
$\theta = 180^{o}$, the $\approx 24\%$ nuclear
depletion of $W(\vec{p}_{m})$ at $p_{m}=0$ goes away with
increasing $p_{m}$ and is superseded by nuclear enhancement
at $p_{m}\gsim 1\,fm^{-1}$. In parallel kinematics, at $\theta
=0^{0}$, a similar transition from nuclear depletion to nuclear
enhancement takes place at $p_{m} \gsim 1.5\, fm^{-1}$.
This difference between the parallel and antiparallel kinematics
is due to the before-mentioned forward-backward asymmetry
generated by the nonvanishing value of $\rho$.
The minimum or shoulder-like irregularity
that is present in PWIA at $p_m \approx 1.7 fm^{-1}$, and which
in transverse kinematics is shifted by FSI to $p_m
\approx 1.4 fm^{-1}$, is washed out by the FSI in
longitudinal
kinematics, for both parallel and antiparallel configurations.
Now, $W(\vec p_m)$ is decreasing monotonously.

The results for parallel kinematics, $\theta = 0^o$, show that the
PWIA distribution $N(p_m)$ and the FSI-modified distribution $W(\vec
p_m)$ are very close to each other for missing momenta $p_m > 2
fm^{-1}$, although $W(\vec p_m)$ slightly undershoots the PWIA values.
Must this be taken as evidence that FSI is unimportant in parallel
kinematics, at $\theta = 0^o$? The results shown in
Figs.~\ref{figmdcpa1}b,c do clearly demonstrate this is not the
case. In parallel kinematics, Fig.~\ref{figmdcpa1}b, $W(\vec{p}_m)$
hardly changes from the mean field, $C_{0} = 0$, (dotted line) to the
soft core correlation, $C_o = 0.5$, (dashed line) to the hard core
correlation, $C_{0}= 1$, (solid line), which must be contrasted to a
dramatic sensitivity of the SPMD to the correlation strength $C_{0}$
in the same region of large $p_{m}$.  In antiparallel kinematics, for
$\theta = 180^{o}$, the FSI-modified distribution $W(\vec{p}_m)$ of
Fig.~\ref{figmdcpa1} exhibits a stronger sensitivity to short range
correlations. However, this dependence on the correlation strength
$C_{0}$ is quite counterintuitive, as $W(p_{m})$ decreases
substantially when short range correlations are switched on, remaining
lower than both the PWIA distribution $N(p_{m})$ and the FSI-modified
distribution $W(\vec{p}_{m})$ evaluated in the mean field
approximation $C_{0}=0$. Notice also, that the difference between the
cases of the soft core and hard core correlations in
Fig~\ref{figmdcpa1} is much smaller than the difference between the
PWIA distribution $N(\vec{p}_{m})$ and the FSI-modified distribution
$W(\vec{p}_{m})$ for $C_{0}=1$ in Fig~\ref{figmdwn}.  The results for
the
larger correlation radius $r_{c}=0.6\,fm$, shown in
Fig.\ref{figmdcpa2}, are not any different form those for $r_{c}=0.5$,
the change of $W(\vec p_m)$ from $r_c = 0.5 fm$ to $r_c = 0.6 fm$ is
marginal.

What actually happens in parallel and anti-parallel
kinematics is a manifestation of still another strong
ISC-FSI correlation effect, which in this case is connected
with the real part of the $pN$ elastic scattering amplitude.
In Fig.\ref{figpwplus}  we show separately the large-$p_{m}$
behaviour of
$W_{+}(p_{m})={1\over 2}[W(\theta=0^{o};p_{m})+W(\theta=180^{o})]$,
which is free of the FSI contribution linear in $\rho$  (the terms
$\propto \rho^2$ that are present in $W_+$ are very small). In the
mean field approximation,  $C_{0}=0$,
the large-$p_{m}$ tail of $W_{+}(p_{m})$ is entirely due to the
$\theta$-function effects. As it was discussed in detail in
\cite{Helium4,Deuter,NSZ}, this tail is suppressed by the
correlation effects,
which naturally smoothes out the idealized $\theta$-function
in the Glauber operator which assumes idealized pointlike nucleons.
On the other hand, the PWIA distribution at $C_{0}=1$
is numerically very
close to the FSI effect at $C_{0}=0$. When both the FSI and
ISC effects are included, with the increase of $C_{0}$ the
rising correlation contribution compensates partly for a decrease
of the FSI contribution. The net effect is a weak depletion
of $W_{+}(p_{m})$ from the mean field, $C_{0}=0$, value to
the soft core correlation value at $C_{0}=0.5$. However,
there is hardly any change in $W_{+}(p_{m})$ from the soft
core to hard core correlation result for $W_{+}(p_{m})$. This is
due to a certain numerical conspiracy between the correlation
and FSI parameters. From the practical point of view it is
important that for a weak energy dependence of the $pN$
scattering parameters, such a conspiracy will persist over
the whole range of $Q^{2}$ to be explored at CEBAF.

The substantial r\^ole of the real part of the $pN$ scattering
amplitude in this region of large $p_{m}$ is obvious from the
FSI-induced forward-backward asymmetry
\begin{equation}
A_{FB}(p_{m})={W(\theta=0^{o};p_{m})-W(\theta=180^{o};p_{m})\over
2W_{+}(p_{m})}\, ,
\label{deffba}
\end{equation}
which is shown in Fig.\ref{figfba}. It exhibits a strong dependence
on the
correlation strength and is quite large for the hard core correlation
$C_{0}=1$.
The forward-backward asymmetry is an intricate
FSI-PWIA interference
effect proportional to the real part of the $pN$ scattering
amplitude.
In the absence of short range correlations, $C_{0}=0$,
the asymmetry stays negative valued at all $p_{m}$. It changes
sign when correlation effects are included. In the latter case
we can compare our results for $^4He$ with the results for the
deuteron, which should resemble each other.
Indeed, for the soft
and hard core correlations, both the magnitude and $p_{m}$
dependence of the asymmetry shown in Fig.~\ref{figfba} are remarkably
similar
to the forward backward asymmetry in $D(e,e'p)$ scattering [10].
For the deuteron target, the calculations are performed using
the realistic wave functions which directly
include the effects of short
distance proton-neutron interaction. From this comparison
we can conclude that, first, our simple Ansatz wave function
correctly models gross features of short-distance nucleon-nucleon
interaction in the $^4He$ and, second, the found change of
the sign of $A_{FB}(p_{m})$ and its rise with the correlation
strength at large $p_{m}$ are on firm grounds.
It is this enhancement of $A_{FB}(p_{m})$
which effectively cancels the effect of slight decrease of
$W_{+}(p_{m})$ and produces the correlation independent
$W(\theta=0^{o};p_{m})$.
It is this enhancement of $A_{FB}(p_{m})$ which amplifies
the slight decrease of $W_{+}(p_{m})$ and produces the
counterintuitive substantial decrease of $W(\theta=180^{o},p_{m})$
with the correlation strength $C_{0}$. We checked that the
(in)sensitivity of $W(\theta=0^{o},180^{o};{p}_{m})$ to the
correlation strength $C_{0}$ decreases with the value of
$\rho$. Having established the origin of masking effects of FSI
on $W(\vec{p}_{m})$ in the longitudinal kinematics, we wish
to point out that in the $^{4}He(e,e'p)$
reaction there is one spectator proton but there are two
spectator neutrons.
The direct experimental knowledge of $\rho$ for the $pn$
scattering is marginal; in our estimates of FSI effects we
rely upon the dispersion theory calculations reviewed in
\cite{lll93}. Even if the uncertainties with the value of
$\rho$ can be eliminated by accurate measurements of
$W_{+}(p_{m})$, the sensitivity of this quantity to the
value of $C_{0}$ is not sufficiently strong for a reliable
separation of the ISC component of $N(p_{m})$.

The overall conclusion from the above discussion is that, despite
the FSI effects in the longitudinal kinematics being less striking
than in the transverse kinematics, in no part of the phase
space is an unambiguous extraction of the short range correlation
component of the
SPMD $N(\vec{p}_{m})$
from the experimental data on missing momentum distribution
$W(\vec{p}_{m})$ possible.




\section{On the convergence of the power expansion in ISC and FSI}

In section 3 we already commented on the importance of higher order
effects in the two-body correlation function
$C(\vec{r}_{i}-\vec{r}_{j})$.
Roughly, the important
terms of each order differ by a factor of $ \left( \frac{r_c }{R_o}
\right)^3 $, i.e. by $0.6$ for $r_c = 0.5 fm$ and by $0.1$ for
$r_c = 0.6 fm$. (Indeed, as Figs.~\ref{figpa1cg},\ref{figpa2cg} show,
 the rate of  convergence
for $r_c = 0.6 fm$ is visibly slower than for $r_c = 0.5 fm$.)
However, there are also large combinatorial factors, which make
the higher order terms non-negligible. In the
PWIA case, the highest existent order is twelve, and the number of
terms in the $n$-th order is $ {12} \choose{n} $, so we
have $66$ second order terms compared to 495 terms of fourth order
to $924$ terms of sixth order
and still $495$ terms of eighth order. This is a reason why
the contribution of sixth order terms to $N(p_{m})$ is still
non-negligible. As a general rule, for a specific
term to be important at high $p_m$, it
has to connect the two trajectories of the OBDM with correlations, i.e.
it has to contain an interaction of the type
$C^{\dagger}(\vec{r}_{4}-\vec{r}_{i})C(\vec{r}_{4}\,'-\vec{r}_{i}\,')$
(plus any number of other correlations between the spectator nucleons
and/or the spectator and struck nucleon trajectories) and/or
a cyclic chain of correlations , e.g.
$C(\vec{r}_{4}-\vec{r}_{i}) C(\vec{r}_{i}-\vec{r}_{j})
C(\vec{r}_{4}\,'-\vec{r}_{j}\,')$. To the leading order, only
three of the possible 66
$C^{\dagger}C, CC, C^{\dagger}C^{\dagger}$
 terms contribute
at large $p_{m}$, and the
percentage of important terms per order increases with the order.
The above points are illustrated by
a comparison of specific higher terms (\ref{eq:4.9})
and (\ref{eq:4.10}) with the leading term (\ref{eq:4.8}).
The importance of higher order effects
is still more enhanced when FSI is included. Here, the
expansion (\ref{eq:4.4}) contains $2^{18}$ terms.

{}From
the practical point of view, the major complication with FSI is that
the longitudinal coordinate integrations and the corresponding Fourier
transforms have to be carried out numerically. At $p_{m,z}=0$, though,
the r\^ole of the correlation and FSI terms in the expansion
(\ref{eq:4.4}) is very similar, and so are the convergence properties.
Here we first present the pure FSI effects in the absence of
correlations, $C_{0}=0$.
Fig.~\ref{figpa4cg}a  shows the convergence of $W(\theta=90^{o},p_{m})$
for
transverse kinematics. In the exact calculation, the momentum
distribution $W(\vec p_m)$ is positive valued, to the lowest odd
orders in $\Gamma^{\dagger}, \Gamma$ one can run into negative valued
$W(\vec{p}_{m})$.
To higher orders in FSI, one introduces attenuation of the
proton wave which leads to a depletion of
$W(\theta=90^{o},p_{m})$. Although to the fourth-order approximation
one finds the positive valued, and slowly decreasing,
 contribution (\ref{eq:4.8}) from the
double incoherent elastic rescattering, which could have enhanced
$W(\theta=90^{o};p_{m})$ somewhat, its normalization is too small to
have an appreciable impact on the $p_{\perp}$ distribution in the
considered region of $p_{m}$. The situation in parallel and
anti-parallel kinematics is very similar, see Figs.~\ref{figpa4cg}b,c.
Here the
crucial point is that FSI generates the large-$p_{m}$ tail in
$W(\vec{p}_{m})$ even in the absence of short range correlations. The
convergence is good and the change from the fourth to sixth order is
marginal. In the considered case, the precocious convergence in
(anti)parallel kinematics is due to an absence of correlations,
see below and the discussion in section 5.

The convergence of the expansion in the correlation function, which
was quite slow already in the PWIA, worsens when correlations
are included. To have some idea on the interplay of FSI and ISC
terms, we show how $W(\vec{p}_{m})$ evolves when the expansion
(\ref{eq:4.4}) is truncated at terms $\sim \Gamma^{k}C^{N-k}$.
In Figs.~\ref{figpa1cg},\ref{figpa2cg}, we present the results only
for
even order N, because to odd orders one
can run into negative valued $W(\vec{p}_{m})$.
The results for $N\geq 6$ include the numerically
stronger FSI effects to all orders. For transverse
kinematics, we find a convergence at large $p_{m}$
only for $N\geq 10$.
In parallel and antiparallel kinematics, the contribution from
$N=12$ is still non-negligible.
Here a part of the problem is an unexpectedly strong FSI-ISC
interference effect
associated with the real part of the $pN$ scattering amplitude,
which exhibits a strong sensitivity to short range correlations.
To summarize, these results demonstrate
that estimates of
the missing momentum distribution to lowest orders in the
correlation function and in the quasideuteron approximation
are too crude for making quantitative conclusions.
For the soft core correlation, $C_{0}=0.5$, the
convergence is much faster.




\section{Implications for nuclear transparency studies}

Nuclear effects in $A(e,e'p)$ scattering are often discussed
in terms of the transparency ratio
$$
T_{A}(\vec{p}_{m}) = {W(\vec{p}_{m}) \over N(\vec{p}_{m})} \, .
$$
The above presented results strongly support the  point
that the FSI effects do not reduce
to an overall renormalization of the observed missing momentum
distribution by the nuclear attenuation factor, which is
an often uncritically made assumption.  Only at
$p_{m}\approx 0$ the found $\approx 24\%$ depletion can be
interpreted as a pure nuclear attenuation effect; at larger
missing momenta, a strong interplay of the attenuation and
distortion effects leads to a nuclear transparency
$T_{A}(\vec{p}_{m})$ which exhibits both much stronger
depletion and "antishadowing" behaviour $T_{A}(\vec{p}_{m})
\gg 1$. This is clearly seen in
Fig.~\ref{figtrans}, in which we show nuclear transparency
for transverse, parallel and anti-parallel kinematics for
the hard core correlation.
In the latter two cases, nuclear transparency is very strongly
affected by the real part of the $pN$ scattering amplitude,
the effect of which can not be interpreted in terms of
attenuation altogether.

In the experimental determination of nuclear transparency
one inevitably runs into a sort of vicious circle: What is
measured experimentally is the FSI-distorted missing
momentum distribution $W(\vec{p}_{m})$ and one is forced
to rely upon some model calculations of the PWIA distribution
$N(\vec{p}_{m})$.
(Still further complications and extra model dependence
will be involved
if the experimentally measured cross section does not allow
an integration over a sufficiently broad range of missing
energy $E_{m}$.)
To a certain extent, this model SPMD $N(\vec{p}_{m})$ can be
checked against the experimentally measured missing momentum
distribution $W(\vec{p}_{m})$, implicitly and/or explicitly
assuming that the FSI effects can be factored out as an overall
attenuation factor. The above presented results
(see also \cite{NSZ}) very clearly show this is not the case and
in large parts of the phase space
such a poor man's evaluation of nuclear transparency can lead
astray. Such a procedure was used,
for instance, in an analysis of the data from the recent
NE18 experiment \cite{NE18exp}. Apart from the factorization
assumption, in the NE18 analysis quite a large $p_{m}$-independent
renormalization was applied to the model SPMD
$N(\vec{p}_{m})$ in anticipation of
the short-range correlation induced reshuffling of strength from
the small to large missing momenta. Here we only wish to
observe, that the results of section 3, see Fig. \ref{figrc},
cast a shadow on such
an oversimplified treatment of the correlation effects.

Having performed a full analysis of the combined ISC and FSI
effects and having established a primacy of FSI effects, we
are in the position to address the question of how
strongly nuclear transparency is sensitive to correlation
effects. In Fig.\ref{figtrans} we present a nuclear
transparency calculated for different correlation
strength,
\beq
T_{A}(C_{0};\vec{p}_{m}) =
{W(C_{0};\vec{p}_{m}) \over N(C_{0};\vec{p}_{m})} \, .
\label{eq:7.1}
\endeq
At large missing momenta, $p_{m}\gsim 1.5\,fm^{-1}$, the
transparency becomes very sensitive to the correlation
strength. In the region of moderate missing momenta,
$p_{m} \lsim  1.3 \,fm^{-1}$, though, nuclear transparency
exhibits hardly any sensitivity to the correlation strength,
which confirms the anticipation in \cite{NSZ}.

The experimental data are often presented in terms of a nuclear
transparency ratio for the cross sections integrated over a certain
momentum window. We present our results for
 the longitudinal and perpendicular
partially integrated transparencies $T_L$ and $T_{\perp}$
defined as follows:
\begin{equation}
T_L (p_{\perp}=0,p_z) = \frac{\int_0^{p_z} dp_z' \, W(
p_{\perp} = 0, p_z')}
{\int_0^{p_z} dp_z' \, N(p_{\perp} = 0, p_z')}
\label{deftl}
\end{equation}
\begin{equation}
T_{\perp} (p_{\perp},p_z=0) = \frac{\int_0^{p_{\perp}} dp_{\perp}'
\, p_{\perp}' \, W(p_{\perp}', p_z=0)}
{\int_0^{p_{\perp}} dp_{\perp}' \, p_{\perp}' \, N(p_{
\perp}', p_z = 0)}
\label{deftperp}
\end{equation}
and the integrated transparency $T_{int}$
\begin{equation}
T_{int}(p_{\perp}) = \frac{\int_0^{p_{\perp}} dp_{\perp}'
 \, p_{\perp}' \, \int_{-\infty}^{\infty} dp_z \,  W(p_{\perp}',
 p_z)}
{\int_0^{p_{\perp}} dp_{\perp}' \, p_{\perp}' \,
\int_{- \infty}^{\infty} dp_z \,  N(p_{\perp}', p_z)} \, .
\label{deftint}
\end{equation}
In agreement with the discussion in \cite{bffps} the integrated
transparencies
are larger when correlation effects are included. However, the overall
effect is very small confirming the conclusions of Ref. \cite{nsswzz}.

The size of the partially integrated transparencies $T_{\perp}
(p_{\perp})$ and $T_L (p_z)$ with the upper limit of integration
$p_{\perp}$ ($p_z$) is quite natural because the larger $p_{\perp}$
($p_z$), the larger fraction of struck protons deflected by elastic
scattering is included.  What is not so obvious is that $T_{int}
(p_{\perp} \sim 0)$ is not really smaller than the fully integrated
transparency $T_{int} (\infty)$.  The naive semiclassical expectation
is that $1 - T_{int}(p_{\perp} \sim 0) \propto \sigma_{tot}(pN)$
vs. $1 - T_{int} (\infty) \propto \sigma_{in} (pN)$, because in
counterdistinction to the former case in the latter case the deflection
of struck protons by elastic rescatterings must not contribute to
nuclear attenuation. This expectation is not born out by exact
calculation, which
demonstrates the pitfalls of the semiclassical
treatment of $W(\vec p_m)$ and the importance of the point that FSI
distortions do not admit the classical interpretation.




\section{Discussion of the results and conclusions}

Quasielastic $A(e,e'p)$ scattering at large missing momentum $p_{m}$
is a natural place to look at the large-$p_{m}$ component of the
single-particle momentum distribution, which is expected to be
generated by short range correlations of nucleons in the ground state
of a target nucleus and which is well known to rise with the
correlation strength.  Our principal finding is that the large-$p_{m}$
behavior of the observed missing momentum distribution in
$^4He(e,e'p)$ is dominated by final state interaction of the struck
proton with spectator nucleons and by the intricate interplay and
quantal interference of the FSI and ground state correlation effects.
In transverse kinematics, the FSI contribution to $W(\vec{p}_{m})$
exceeds the ISC contribution to the SPMD by the order of
magnitude. Even here, a substantial part of the FSI effect comes from
a quantum mechanical FSI-ISC interference effect in the one body
density matrix, which defies a semiclassical interpretation. The
pattern of FSI-ISC interference effects is still more complex for
longitudinal kinematics, where we found a novel effect of strong
enhancement of the forward-backward asymmetry by short range
correlations in the ground state. In antiparallel kinematics, this
ISC-FSI interference effect in the contribution of the real part of
the $pN$ scattering amplitude leads to the FSI-modified
$W(\vec{p}_{m})$ which decreases with the correlation strength in the
opposite to the SPMD.  We are led to the conclusion that FSI effects
make impossible a model-independent determination of the SPMD
$N(\vec{p}_{m})$ from the experimentally measured missing momentum
distribution $W(\vec{p}_{m})$.  Large FSI effects are of quite a
generic origin and are not an artifact of the Ansatz wave function
used in our evaluations, realistic though it is.  We emphasize a
simple and well understood origin of large enhancement parameters
(\ref{eq:4.6},\ref{eq:4.9}), which is a large radius of the nucleus as
compared to a small radius of short range correlations.  There remains
the intriguing possibility of the forward-backward asymmetry as a
probe of short range correlations and further studies of the model
dependence of this observable are worth while.

Our Ansatz for the wave function was motivated by a desire to
have a complete calculation of both short range correlation
and final state interaction effects, rather than an evaluation
of several lowest order terms. Even though much of the integrations
and Fourier transforms can be performed analytically, the
numerical calculations present quite a formidable task.
The above presented results
show that higher order correlation and FSI effects are indeed
important. Our simple Jastrow function only includes the
$S$-wave correlations. Our
previous work on the missing momentum distribution for
$^{2}H(e,e'p)$ scattering suggests that the $D$-wave effects are
not that important \cite{Deuter,Tensor}. Specifically, it has been
shown that the sensitivity to
different models for the deuteron wave function, which give the
$D$-wave contributions to $N(\vec{p}_{m})$ differing at large
$p_{m}$ by an order of magnitude, is completely lost
when including FSI.
This insensitivity towards the D-wave is due to the fact that the FSI
operator is short ranged and therefore does not affect the D-wave very
much, as the D-wave is suppressed at small distances by the
centrifugal
barrier. At the higher missing momenta we are interested in,
the missing
momentum distribution was found to be dominated by FSI distortions
of the $S$-wave contribution. Therefore, we can expect that
the D-wave effects would not change our conclusions on the
relative importance of the FSI and ISC effects at large $p_{m}$.
One of the interesting findings is a substantial effect of higher
order terms in the pair correlation function, which clearly
shows an inadequacy of the oversimplified quasi-deuteron model
for the large-$p_{m}$ component of the missing momentum
distribution, both in PWIA and with allowance for FSI.

One interesting implication of a dominance of FSI effects at
large $p_{m}$ in both $^{2}H(e,e'p)$ and $^{4}He(e,e'p)$
scattering is the similarity of missing momentum spectra
(scaled up by the factor $\sim 3$ for the deuteron).
Such a similarity emerges not because of the quasi-deuteron
mechanism in the $^{4}He$, but because of the universality
of final state proton-nucleon interaction in both nuclei
\cite{Univers}.

Strong FSI effects in parallel kinematics also
affect an interpretation of the y-scaling analysis of $(e,e'p)$
scattering in terms of the SPMD.
To the extent that according to the experimental data,
the FSI parameters - the total $p-n$ cross section,
diffraction slope, the ratio $\rho$
of real to imaginary part of the forward elastic
scattering amplitude
 - vary only slightly for $Q^{2}$ of
several $GeV^2$, the FSI dominated $W(\vec{p}_{m})$ also shall
stay approximately $Q^{2}$ independent over the CEBAF range of
$Q^2$. This fact leads to an ``FSI-scaling'' effect that
should not be confused with real y-scaling.

Last but not least it is well known that for the high
density of $^4He$,
the ISC-induced large-$p_m$ tail of the SPMD does not change
substantially
from $^4He$ to heavy nuclei \cite{Ciofi91,Pieper,Co}. On the
other hand,
the FSI effects rise steeply with the mass number. For instance,
the nuclear transparency decreases from $\sim 0.75$ for
$^4He$ down to
$ \sim 0.25$ for $^{197}Au$ \cite{NE18exp}. This suggests
 strongly that
FSI distortions will be still stronger and ISC effects will become
marginal in $A(e,e'p)$ on heavy nuclear targets.
\bigskip\\

{\bf Acknowledgments:} We acknowledge discussions with
S.Boffi, C.Ciofi degli Atti, S.Simula and J. Speth.
This work was supported in part by the
INTAS Grant No. 93-239 and by the Vigoni Program of DAAD (Germany)
and the Conferenza Permanente dei Rettori (Italy).
A.Bianconi thanks J.Speth for the hospitality at IKP, KFA J\"ulich,
and S.Jeschonnek thanks S.Boffi for the hospitality at the
University of Pavia.
\pagebreak\\


{\bf \Large Figure Captions}

\begin{figure}[h]
\caption {Schematic diagrams of several contributions to the
  one-body density matrix containing correlations. a)
  a linear correlation term of the type $C^{\dagger} (\vec r_i
  \,' - \vec r_4 \, ')$,  b)  the quadratic
  contribution $C^{\dagger} (\vec r_i \,' - \vec r_4 \, ') \, C (\vec
  r_i - \vec r_4)$ and  c) a cyclic third order
  contribution of the type $C^{\dagger} (\vec r_4 \,' - \vec r_i \, ')
  \, C (\vec r_i - \vec r_k) \, C (\vec r_k - \vec r_4)$.}
\label{fig1}
\end{figure}
\begin{figure}[h]
\caption
 {The single particle momentum distribution (SPMD)
  for the mean field distribution $C_o = 0$ (dotted curve), for soft
  core correlations $C_o = 0.5$ (dashed curve) and for hard core
  correlations $C_o = 1$ (solid curve). The other parameters are
  $R_o = 1.29 fm$ and $r_c = 0.5 fm$.}
\label{figpwc}
\end{figure}
\begin{figure}[h]
\caption
  {The single particle momentum distribution (SPMD)
  for different values of the correlation radius $r_c$.  The solid
  curve shows $r_c = 0.5fm$ and the dashed curve shows $r_c =
  0.6fm$. The other parameters are $R_o = 1.29 fm$ and $C_o = 1$.}
\label{figpwrc}
\end{figure}
\begin{figure}[h]
\caption
  {The convergence behaviour of the single particle
  momentum distribution for $R_o = 1.29 fm$, $r_c = 0.5 fm$ and
  $C_o = 1$. The dotted line includes all terms up to second order,
  the dashed line includes all terms up to fourth order,
  the dash-dotted
  line shows all terms up to sixth order, the double-dotted
  line includes
  all terms up to eighth order and the solid line shows the
  full calculation,
  i.e. it includes all terms up to twelfth order.}
\label{figpwcon}
\end{figure}
\begin{figure}[h]
\caption
 {The pattern of renormalization of the SPMD for the short
range correlation.}
\label{figrc}
\end{figure}
\begin{figure}[h]
\caption
{Schematic diagrams of several contributions to the one-body density
  matrix modified by final state interactions and correlations. Panel
  a) shows a linear final state interaction term of the type
  $\Gamma^{\dagger}
  (\vec
  r_i \,' - \vec r_4 \, ')$, panel b) shows the most important
  quadratic contribution $\Gamma^{\dagger} (\vec r_i \,' -
  \vec r_4 \, ')
  \, \Gamma (\vec r_i - \vec r_4)$ and panel c) shows a second order
   FSI-ISC interference
  contribution of the type $\Gamma^{\dagger} (\vec r_4 \,' -
  \vec r_i \, ')
   \, C (\vec r_i - \vec r_4)$.}
\label{figdiagfsi}
\end{figure}
\begin{figure}[h]
\caption
 {The angular dependence of the missing momentum distribution $W(\vec
  p_m)$ for different missing momenta $p_m$ (full line). For
  comparison, the corresponding value of the single particle momentum
  distribution $N(p_m)$ is also plotted (dotted line).}
\label{figad}
\end{figure}
\begin{figure}[h]
\caption
 {The missing momentum distribution $W(\vec p_m)$ (full line)
  and the single particle momentum distribution $N(p_m)$ (dashed line)
  for $\theta = 90^o$ (upper panel), $\theta = 0^o$ (middle panel),
  and for $\theta = 180^o$ (lower panel). The parameters for the nuclear
   ground state are $R_o = 1.29 fm$, $r_c = 0.5 fm$ and $C_o = 1$.}
\label{figmdwn}
\end{figure}
\begin{figure}[h]
\caption
 {The missing momentum distribution $W(\vec p_m)$ is plotted for hard
  core correlations, $C_o = 1$, (solid line), soft core correlations,
  $C_o = 0.5$, (dashed line) and no correlations at all, $C_o = 0$
  (dotted line), for different angles $\theta$. The upper panel shows
  the results for $\theta = 90^o$, $\theta = 0^o$ is shown in the
  middle panel and $\theta = 180^o$ is shown in the lower panel.
  The other parameters are $R_o = 1.29 fm$ and $r_c = 0.5 fm$.}
\label{figmdcpa1}
\end{figure}
\begin{figure}[h]
\caption
 {The same as in the previous figure, only with the correlation
  radius $r_c = 0.6 fm$ instead of $r_c = 0.5 fm$ as above.}
\label{figmdcpa2}
\end{figure}
\begin{figure}[h]
\caption
 {The distribution $W_{+}(p_{m})={1\over
  2}[W(\theta=0^{o};p_{m})+W(\theta=180^{o})]$ is shown for hard core
  correlations, $C_o = 1$, (solid line), soft core correlations, $C_o
  = 0.5$, (dashed line) and no correlations at all, $C_o = 0$,(dotted
  line). The other parameters are $R_o = 1.29 fm$ and $r_c = 0.5 fm$.
  For comparison, the single particle momentum distribution for hard
  core correlations, $C_o = 1$, is also shown (dash-dotted line).}
\label{figpwplus}
\end{figure}
\begin{figure}[h]
\caption
 {The forward-backward asymmetry $A_{FB}(p_m)$ as defined in (\protect
 {\ref{deffba}}) is shown for hard core correlations, $C_o = 1$, (solid
  line), soft core correlations, $C_o = 0.5$, (dashed line) and no
  correlations at all, $C_o = 0$, (dotted line). The other ground
  state parameters are $R_o = 1.29 fm$ and $r_c = 0.5 fm$.}
\label{figfba}
\end{figure}
\begin{figure}[h]
\caption
 {The convergence behaviour of the missing momentum distribution
 $W(\vec p_m)$
  without any correlations, i.e. $C_o = 0$, for different angles
  $\theta$.
  The dotted line shows all the terms up to second order, the dashed
  line
  represents all terms up to fourth order, and the full line shows the
  complete result, i.e. all terms up to sixth order.  The upper panel
  shows
  the results for $\theta = 90^o$, $\theta = 0^o$ is shown in the
  middle panel and $\theta = 180^o$ is shown in the lower panel.
  The other parameters are $R_o = 1.29 fm$ and $r_c = 0.5 fm$.}
\label{figpa4cg}
\end{figure}
\begin{figure}[h]
\caption
 {The convergence behaviour of the missing momentum distribution
  $W(\vec p_m)$ with hard core correlations, i.e. $C_o = 1$, for
  different angles $\theta$.  The dotted line shows all the terms up
  to second order, the dashed line represents all terms up to fourth
  order, the dash-dotted line shows all terms up to sixth order, the
  double-dotted line includes all terms up to eighth order, the
  long-dashed line shows the calculation up to the tenth order, the
  dash-double-dotted line represents all terms up to twelfth order,
  and the full line shows the complete result, i.e. all terms up to
  18th order.  The upper panel shows the results for $\theta = 90^o$,
  $\theta = 0^o$ is shown in the middle panel and $\theta = 180^o$ is
  shown in the lower panel.  The other parameters are $R_o = 1.29 fm$
  and $r_c = 0.5 fm$.}
\label{figpa1cg}
\end{figure}
\begin{figure}[h]
\caption
 {The convergence behaviour of the missing momentum distribution
  $W(\vec p_m)$ with soft core correlations, i.e. $C_o = 0.5$, for
  different angles $\theta$.  The dotted line shows all the terms up
  to second order, the dashed line represents all terms up to fourth
  order, the dash-dotted line shows all terms up to sixth order, the
  double-dotted line includes all terms up to eighth order, the
  long-dashed line shows the calculation up to the tenth order, the
  dash-double-dotted line represents all terms up to twelfth order,
  and the full line shows the complete result, i.e. all terms up to
  18th order.  The upper panel shows the results for $\theta = 90^o$,
  $\theta = 0^o$ is shown in the middle panel and $\theta = 180^o$ is
  shown in the lower panel.  The other parameters are $R_o = 1.29 fm$
  and $r_c = 0.5 fm$.}
\label{figpa2cg}
\end{figure}
\begin{figure}[h]
\caption
 {The nuclear transparency $T_{A}(\vec p_m)$ as defined in (\protect
  {\ref{eq:7.1}}) is shown for hard core correlations, $C_o = 1$,
  (solid line), soft core correlations, $C_o = 0.5$, (dashed line) and
  no correlations at all, $C_o = 0$, (dotted line). The upper panel
  shows the results for $\theta = 90^o$, $\theta = 0^o$ is shown in
  the middle panel and $\theta = 180^o$ is shown in the lower panel.
  The other ground
  state parameters are $R_o = 1.29 fm$ and $r_c = 0.5 fm$.}
\label{figtrans}
\end{figure}
\begin{figure}[h]
\caption
 {The ratios $T_L$ ($T_{\perp}$) of the $p_z$ ($p_{\perp}$) integrated
spectral functions as function of the integration limit $p_{max}$ as
defined in
(\protect {\ref{deftl}}),
(\protect {\ref{deftperp}})
 are shown for hard core
correlations, $C_o = 1$, (solid line), soft core correlations, $C_o =
0.5$, (dashed line) and no correlations at all, $C_o = 0$, (dotted
line). The upper panel shows the results for $\theta = 90^o$, $\theta
= 0^o$ is shown in the middle panel and $\theta = 180^o$ is shown in
the lower panel.  The other ground state parameters are $R_o = 1.29
fm$ and $r_c = 0.5 fm$.}
\label{figtlp}
\end{figure}
\begin{figure}[h]
\caption
 {The fully integrated transparency $T_{int}$ as function of the
 integration
limit $p_{\perp,max}$ as defined in (\protect {\ref{deftint}}) are shown
for hard core correlations, $C_o = 1$, (solid line), soft core
correlations, $C_o = 0.5$, (dashed line) and no correlations at all,
$C_o = 0$, (dotted line).  The other ground state parameters are $R_o
= 1.29 fm$ and $r_c = 0.5 fm$.}
\label{figtint}
\end{figure}
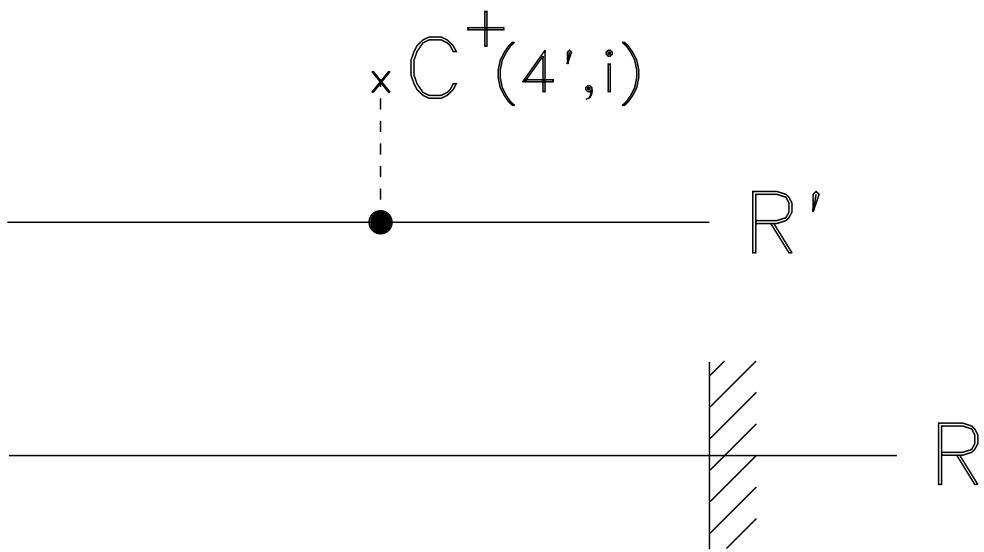
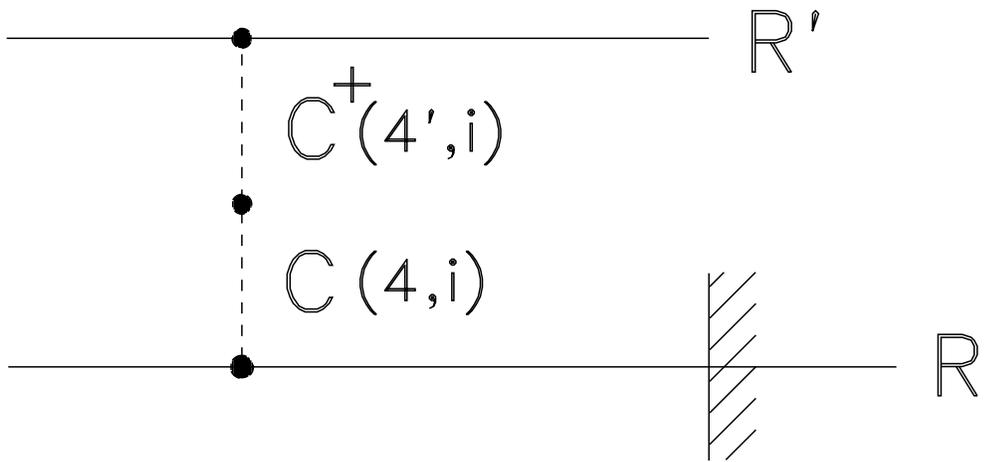
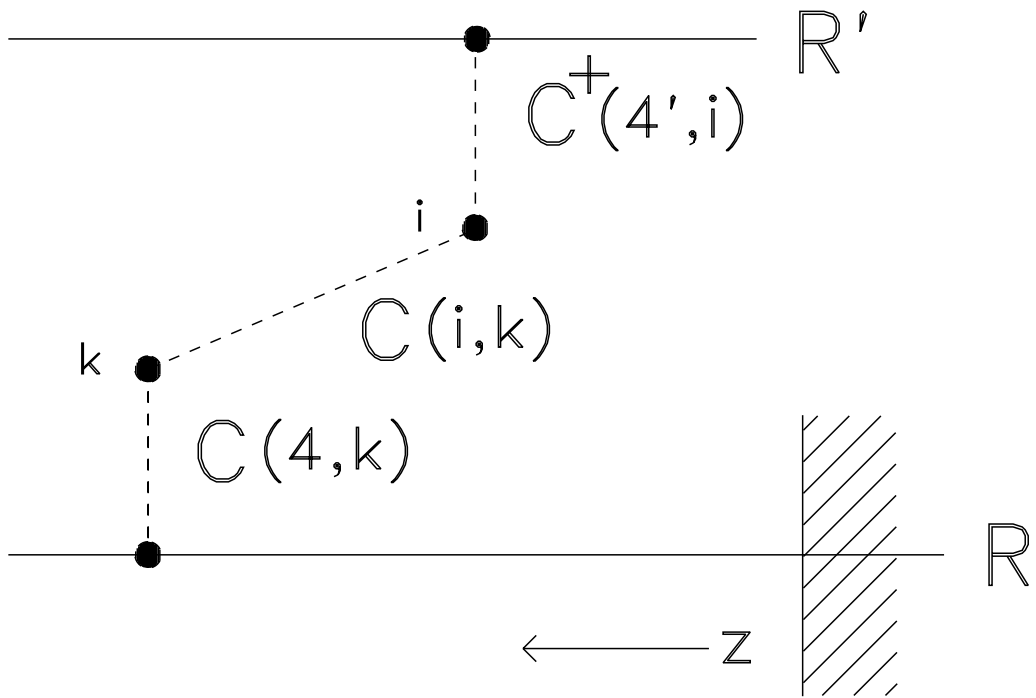
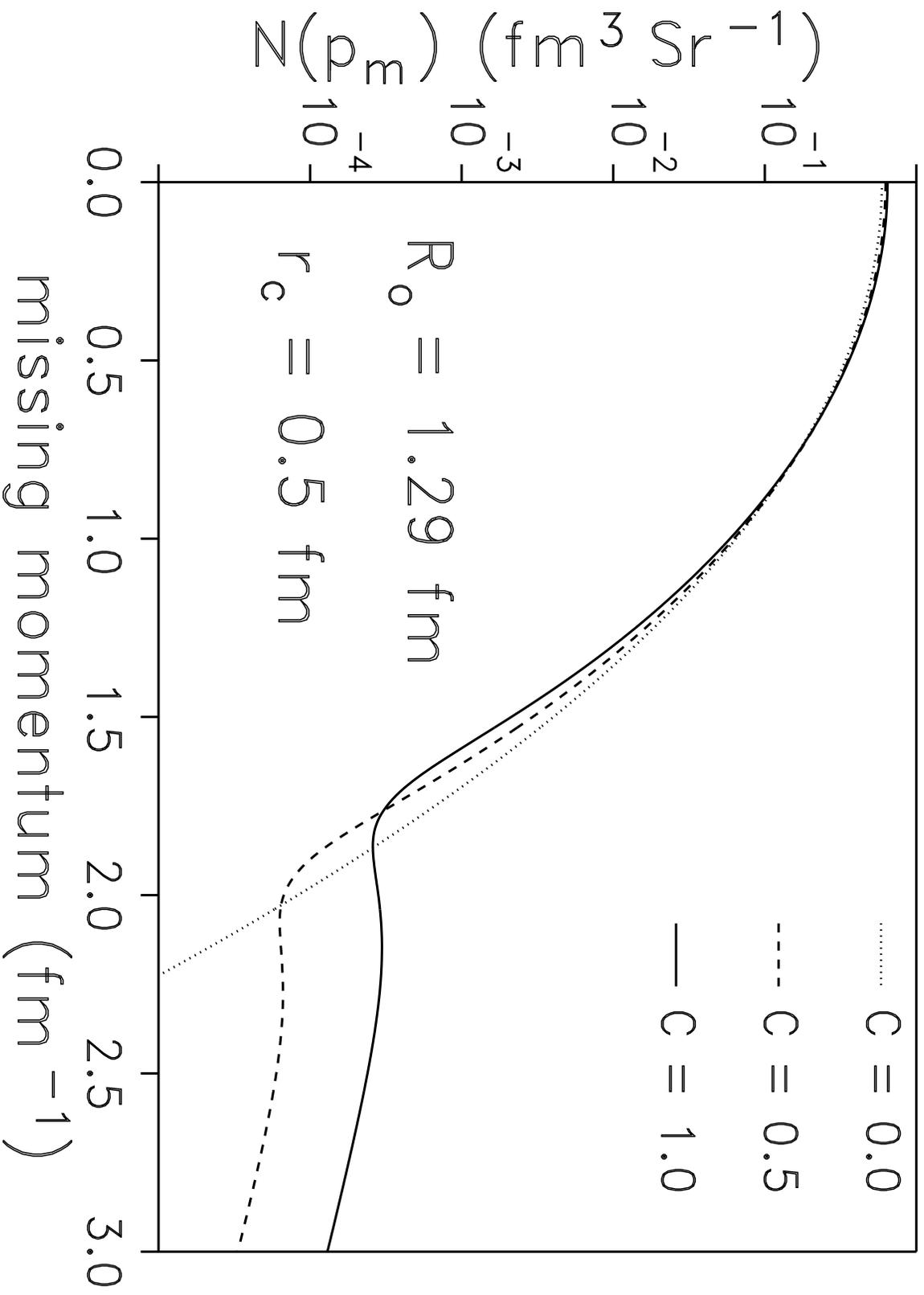
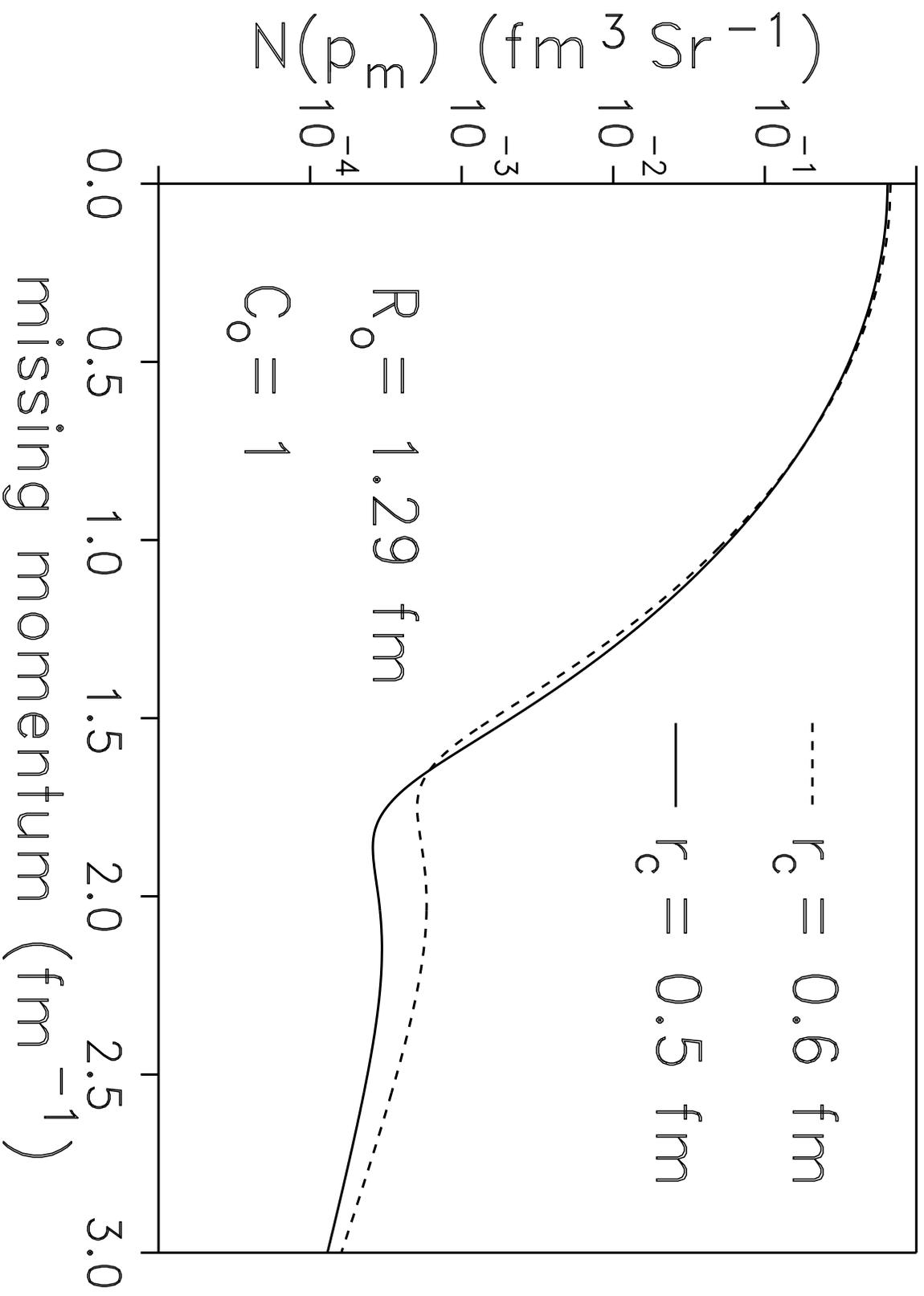
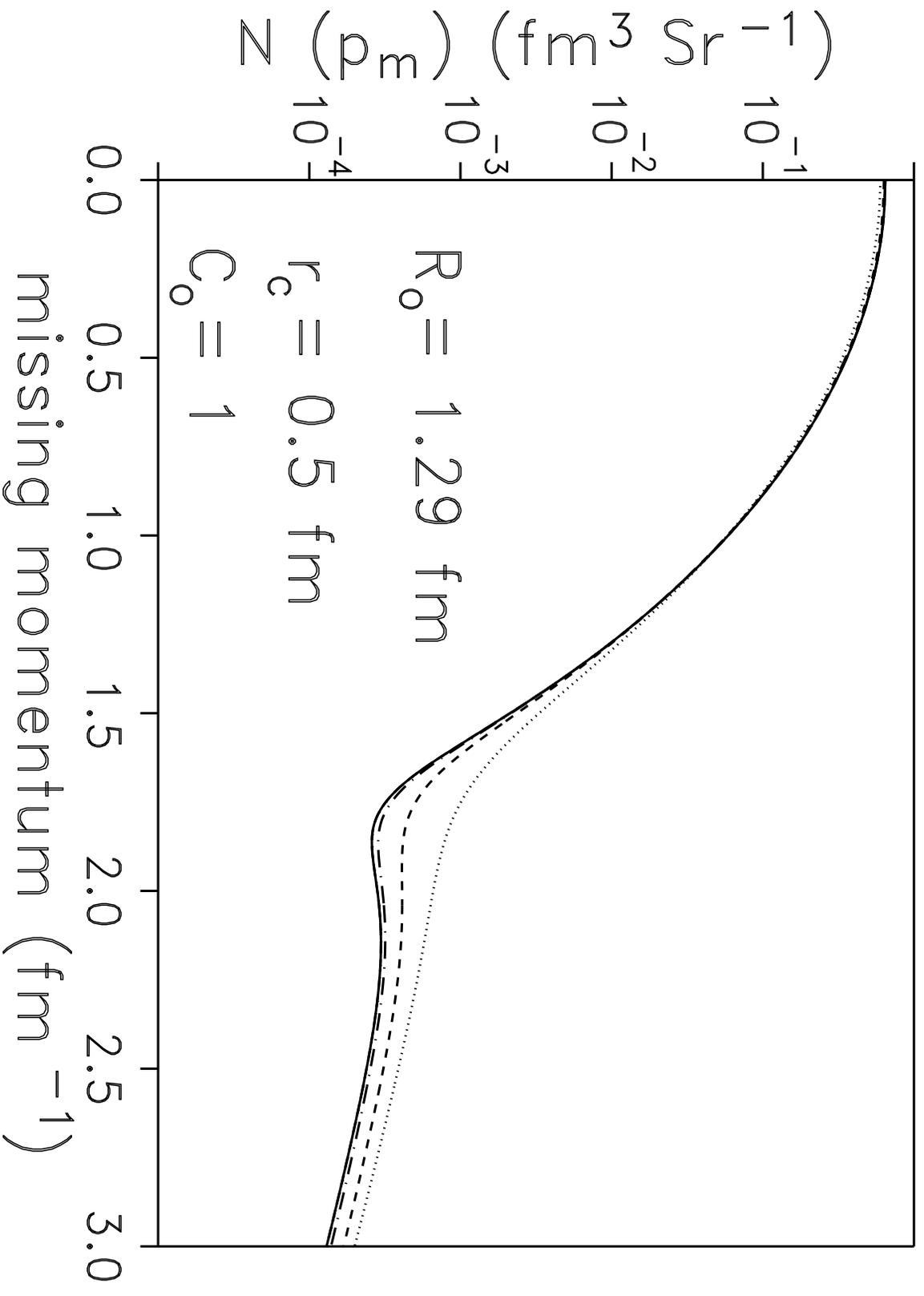
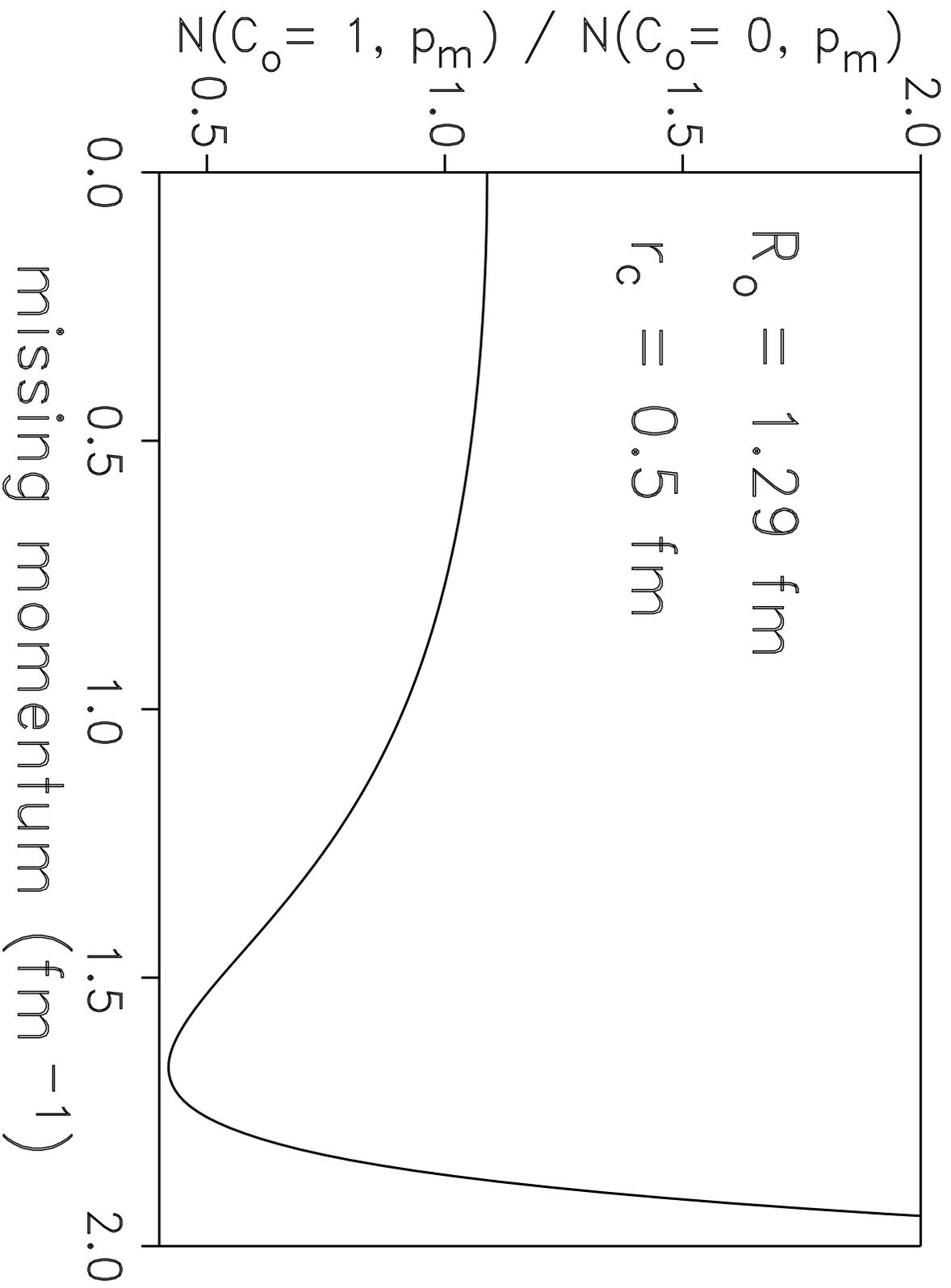
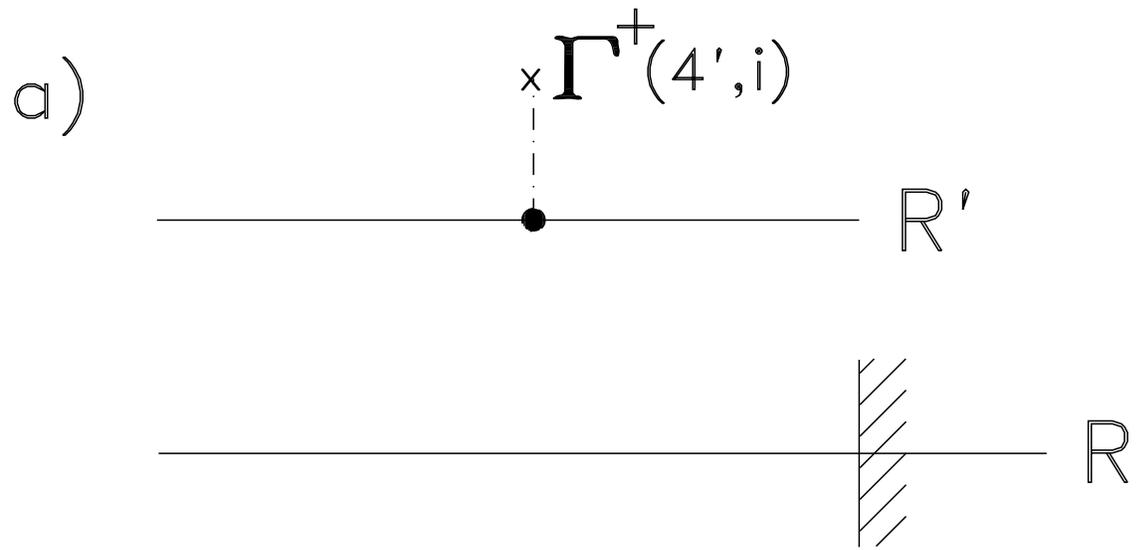
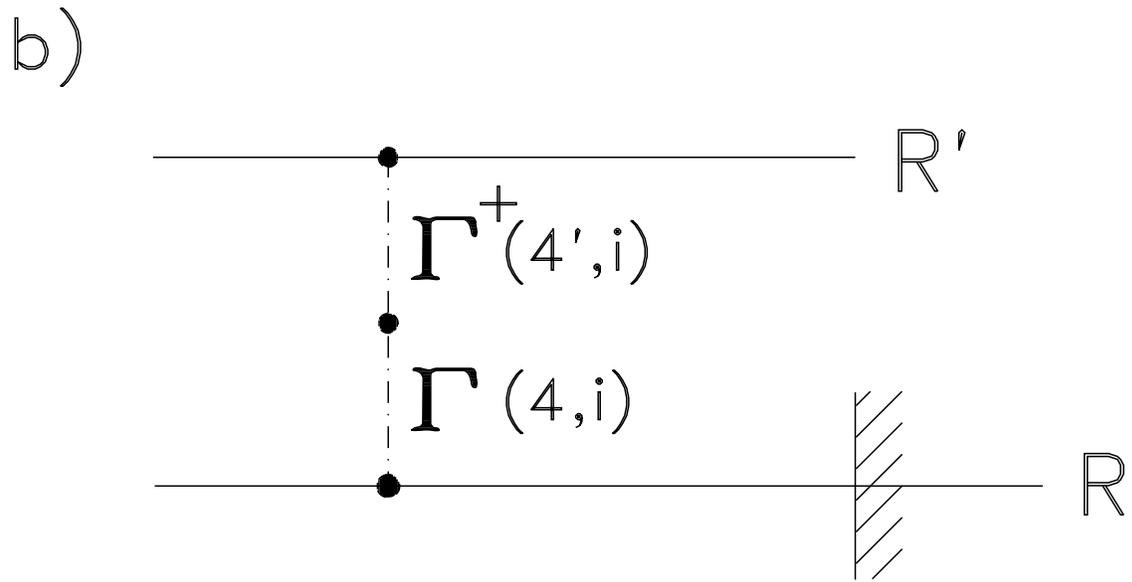
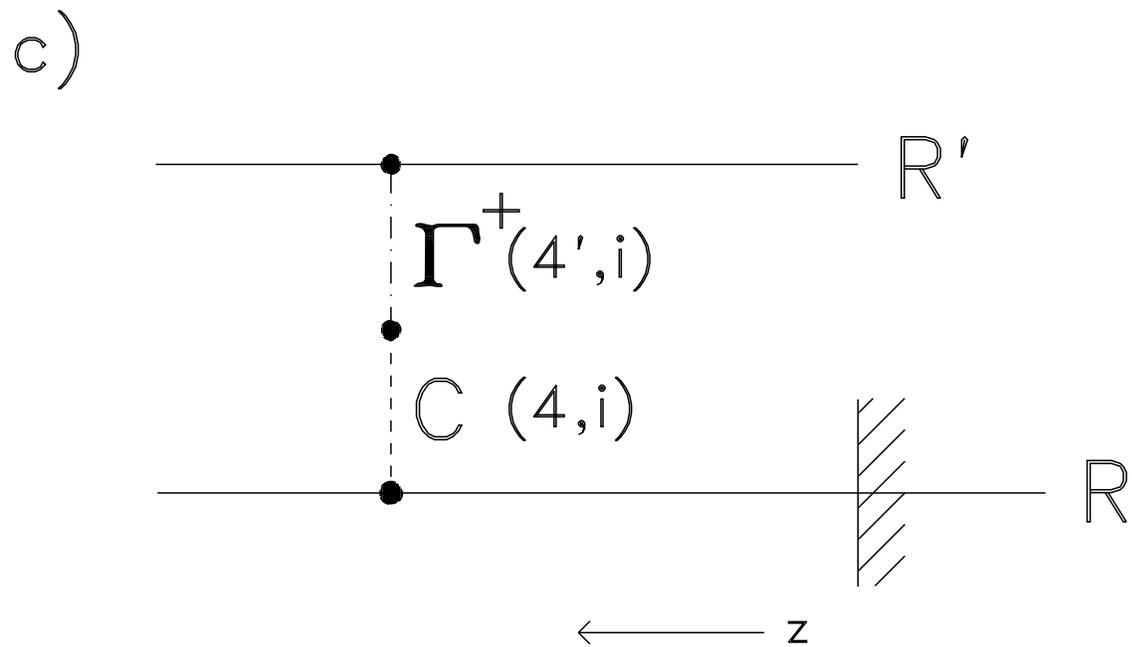
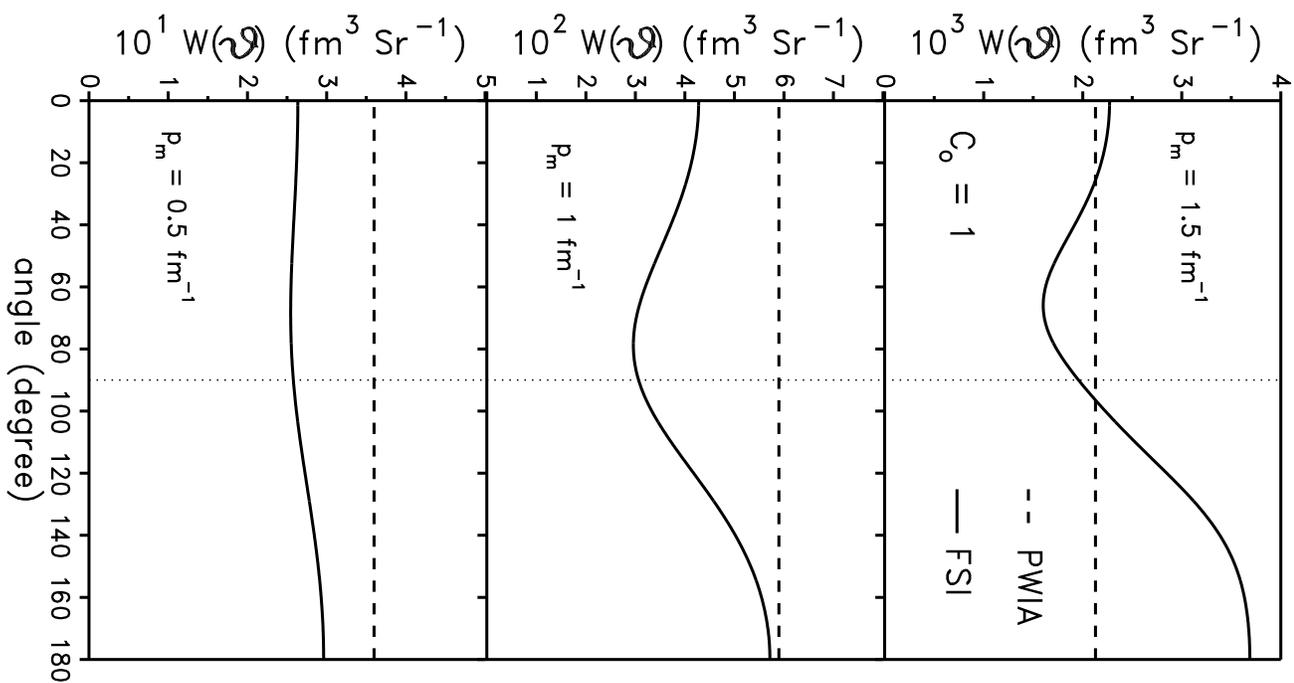
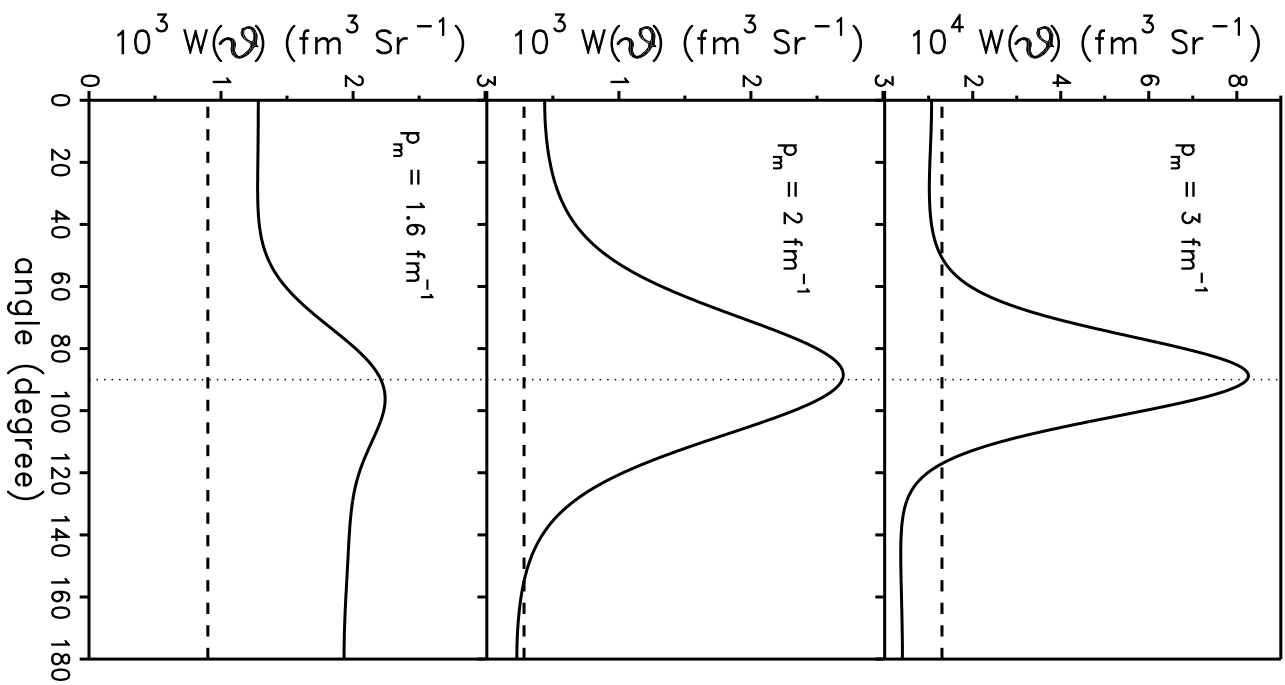
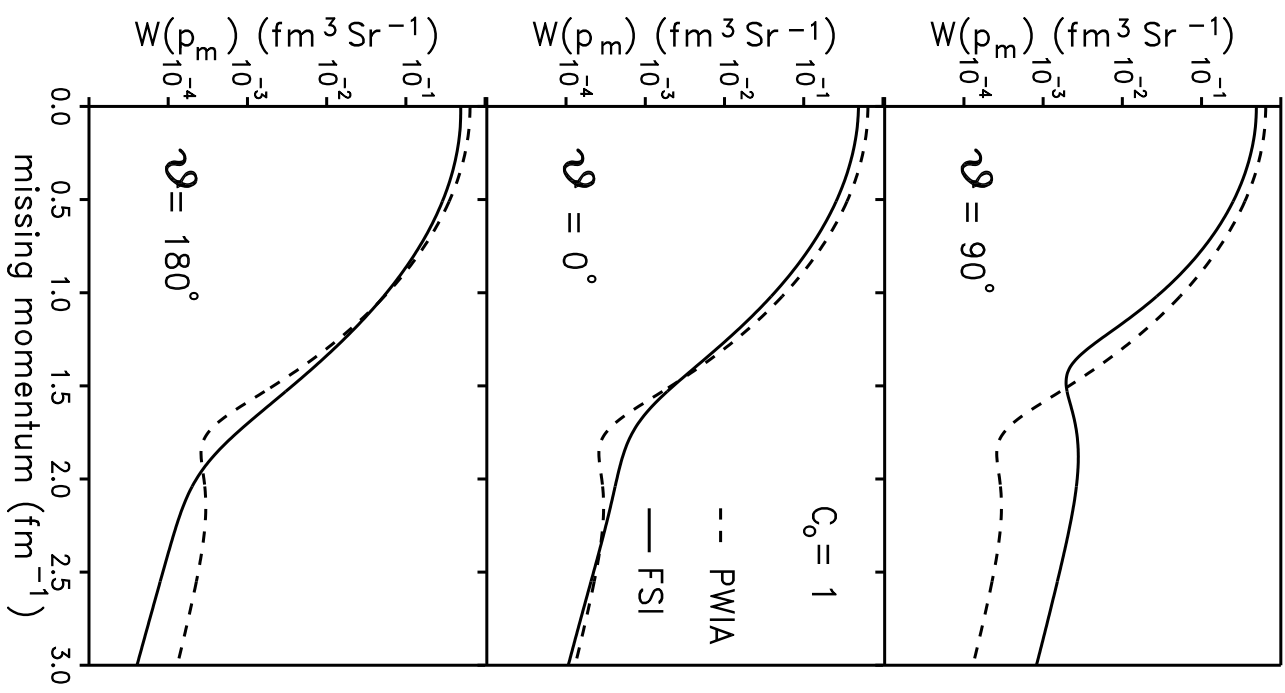
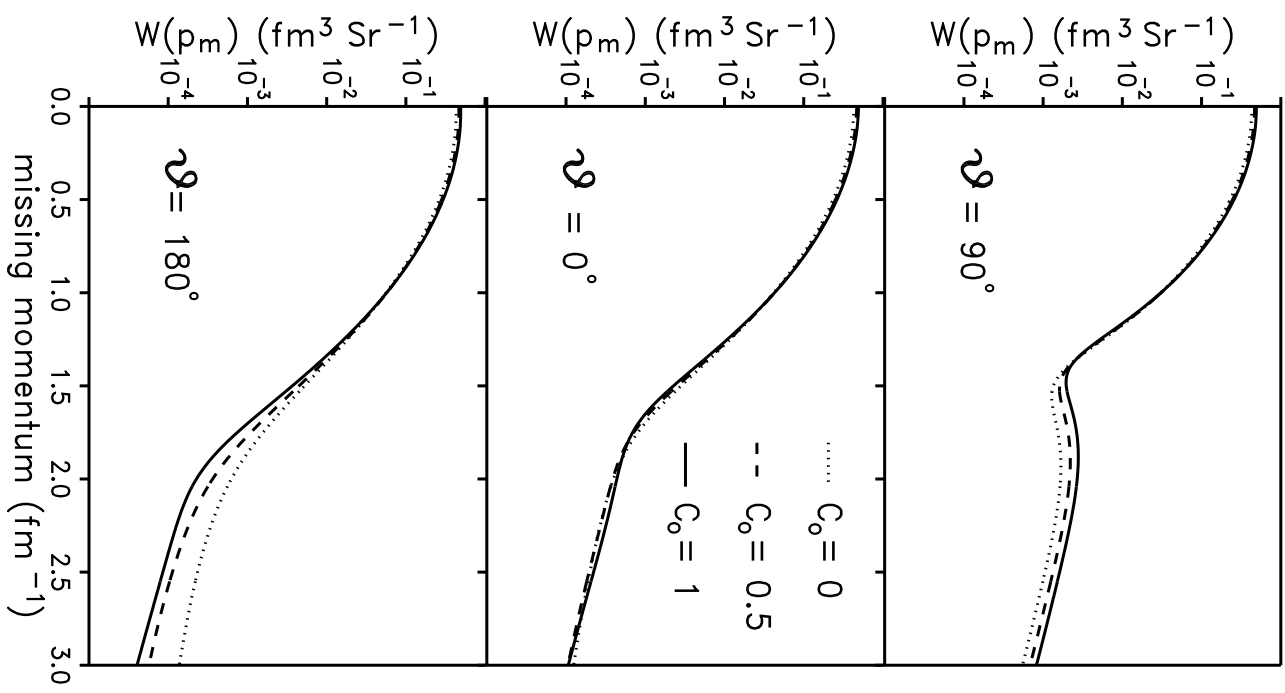
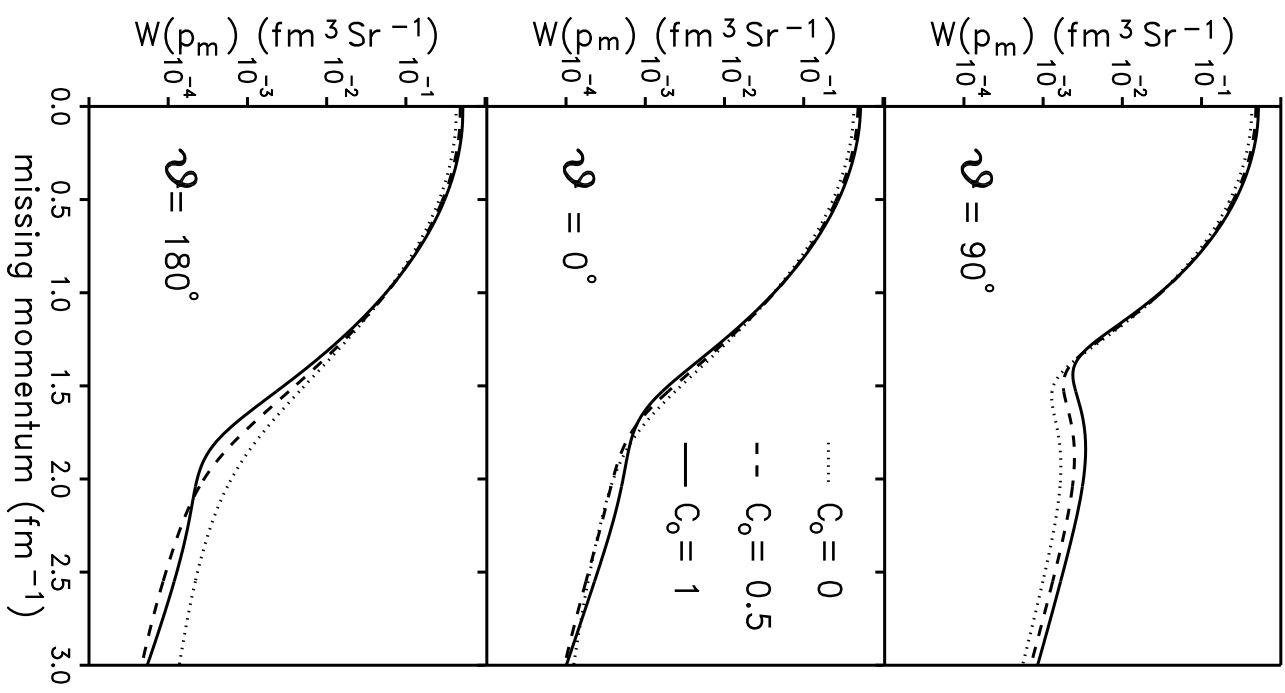
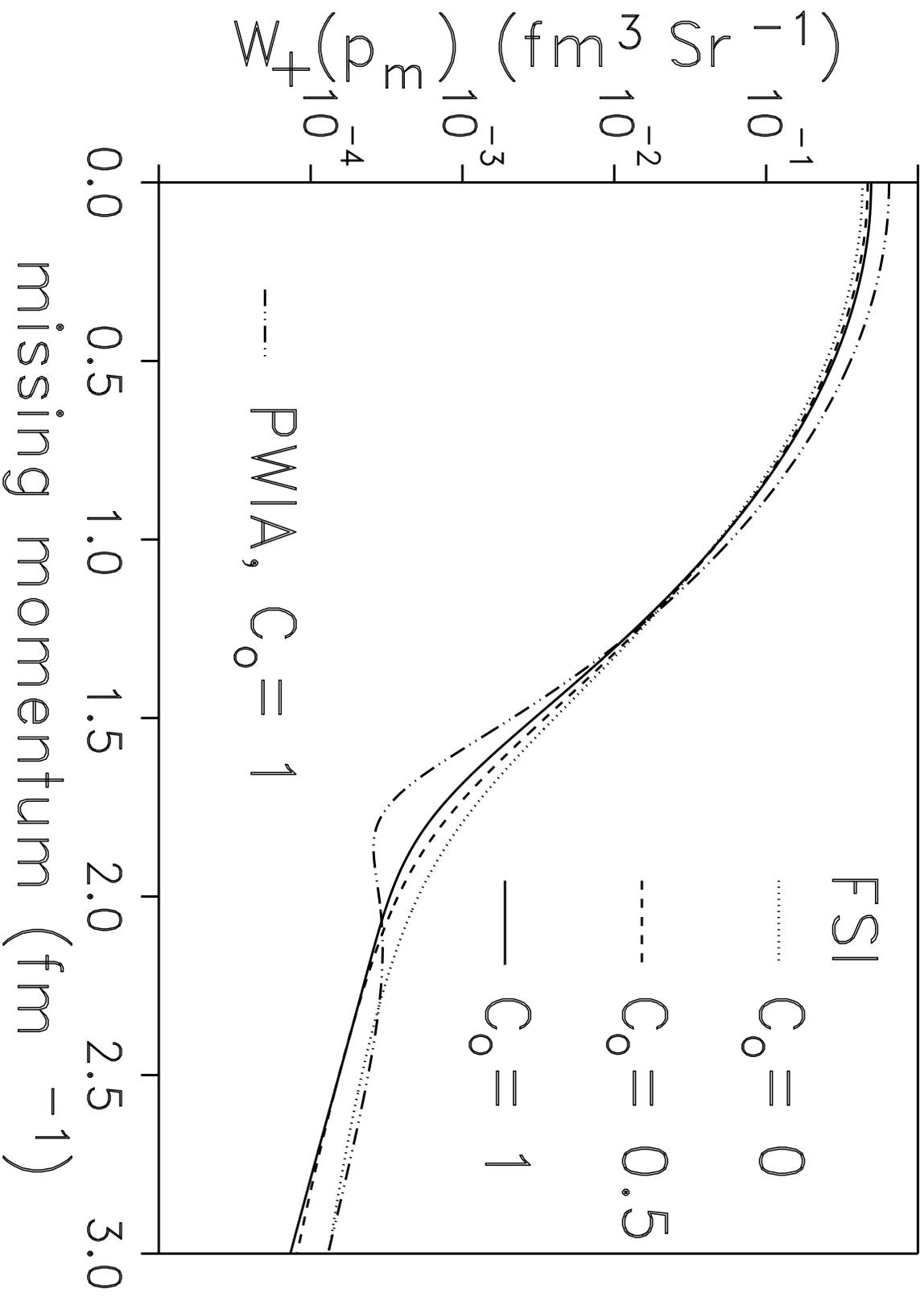
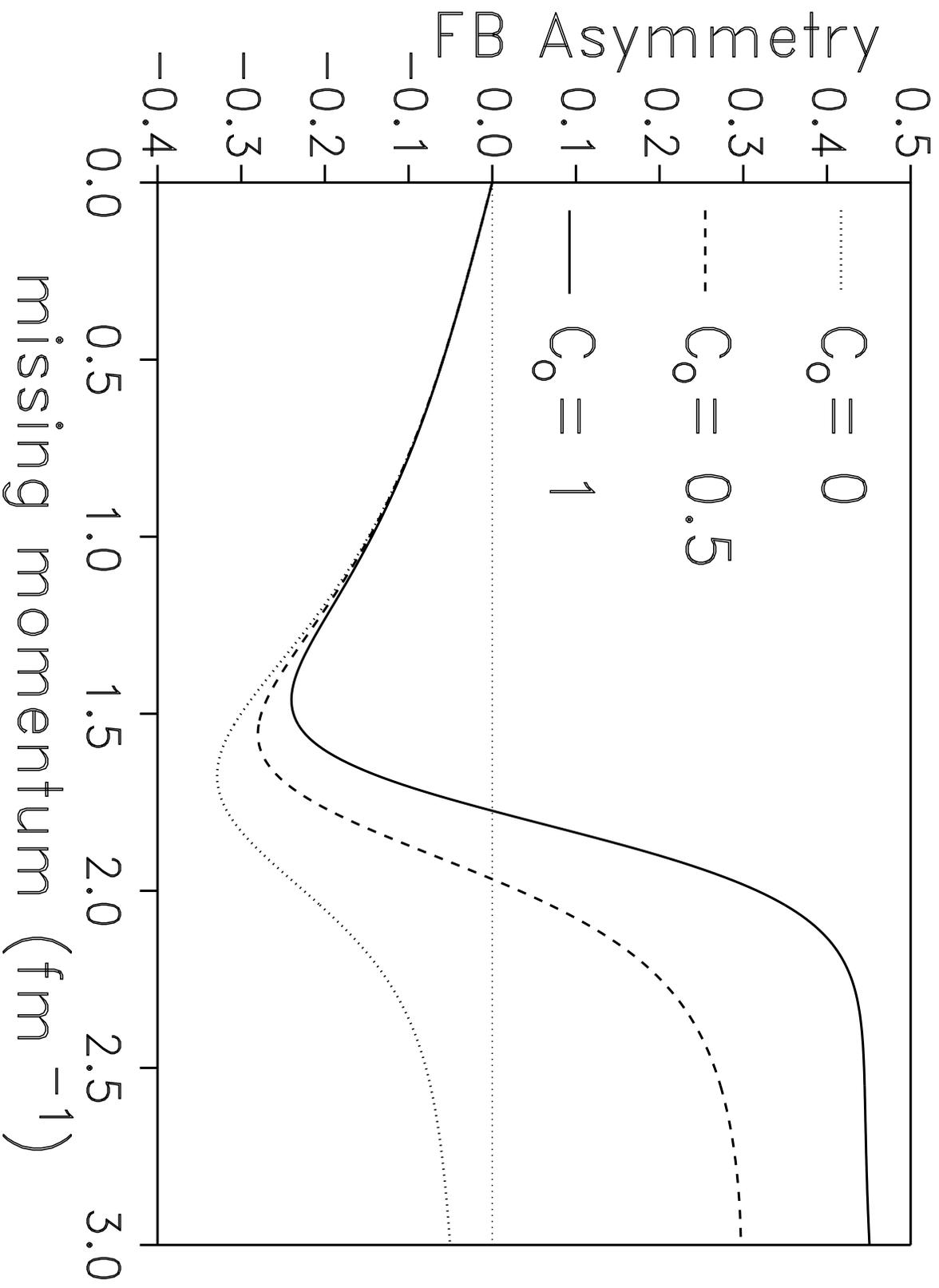
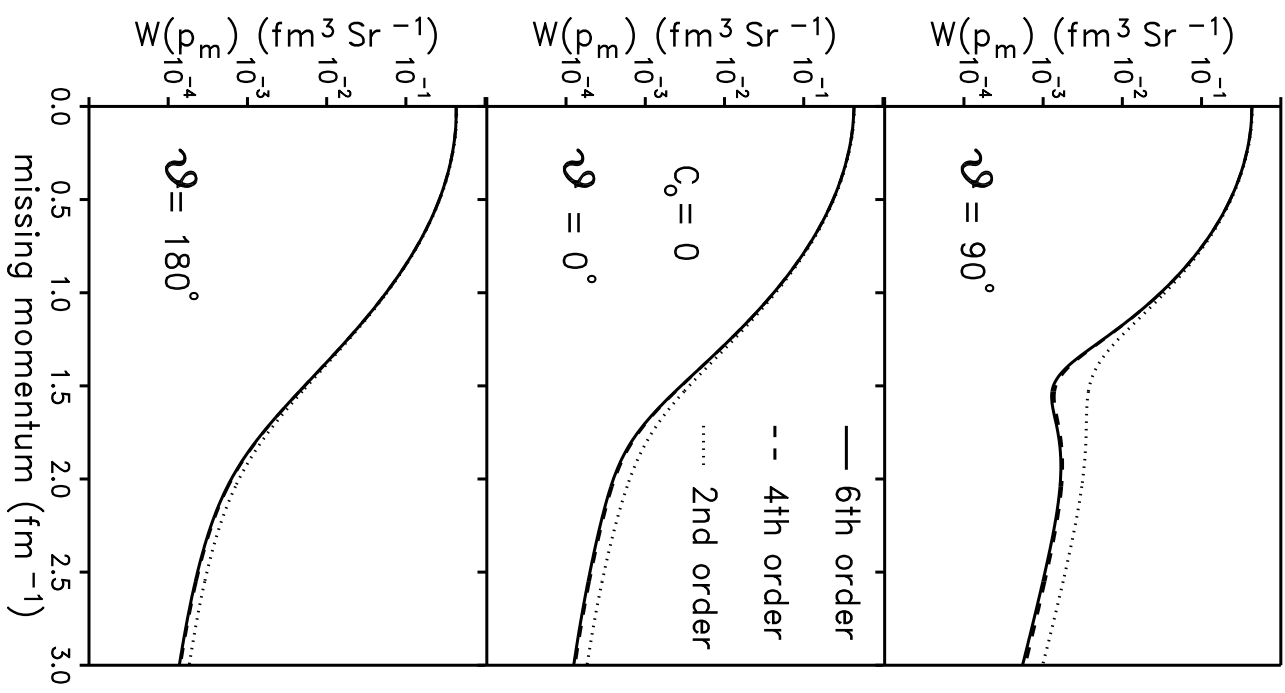
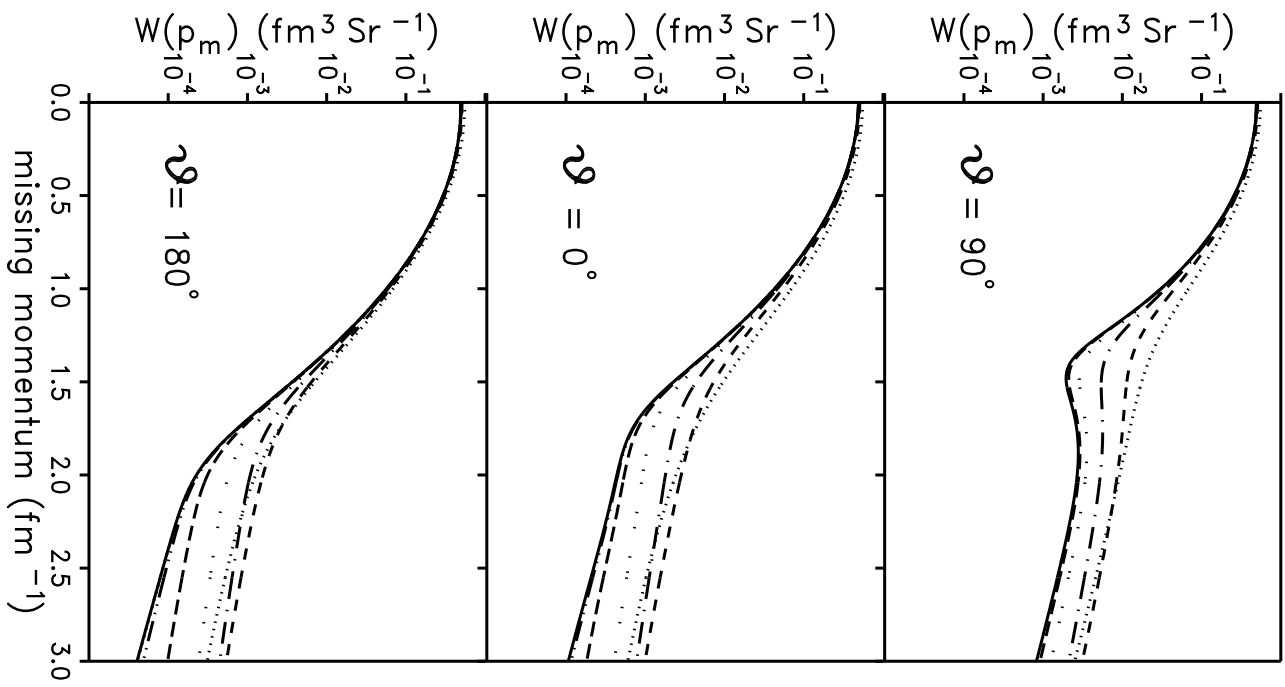
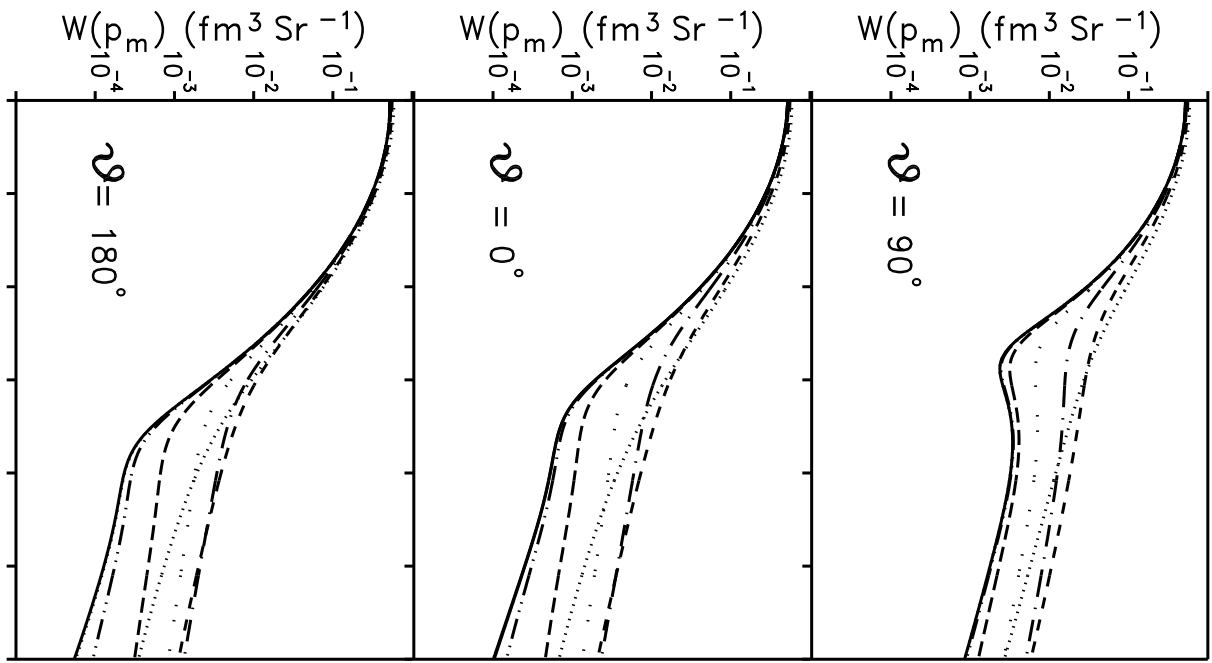
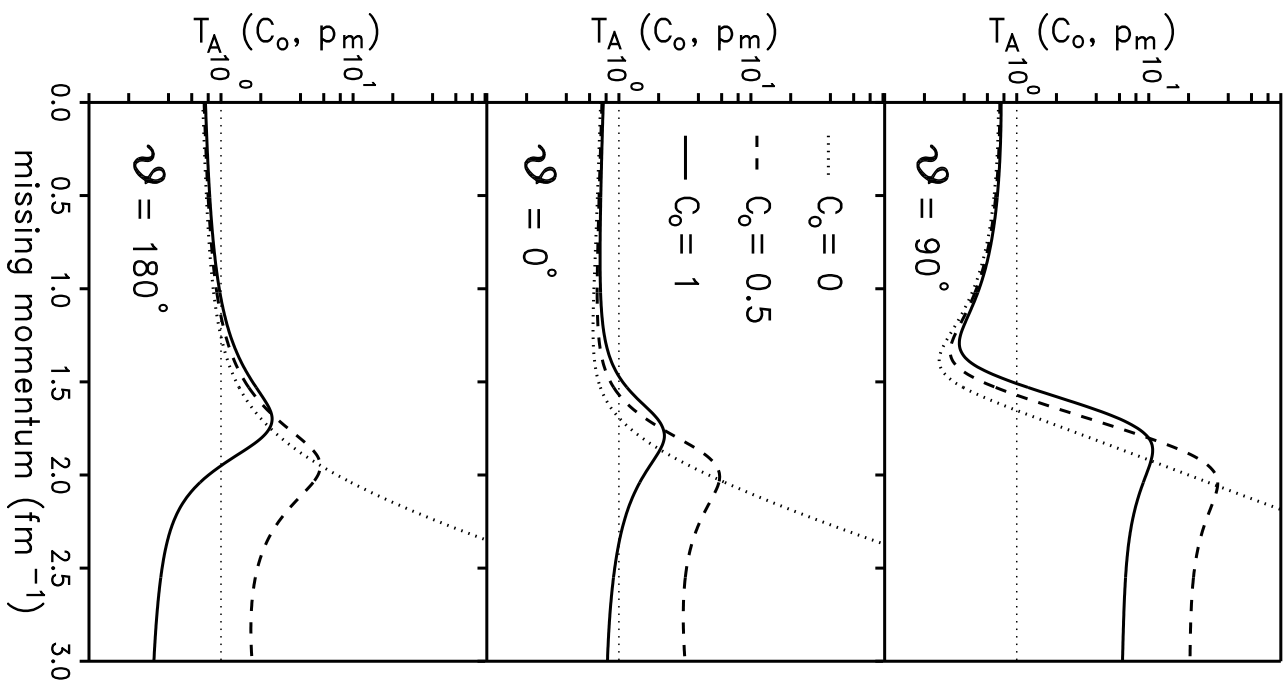
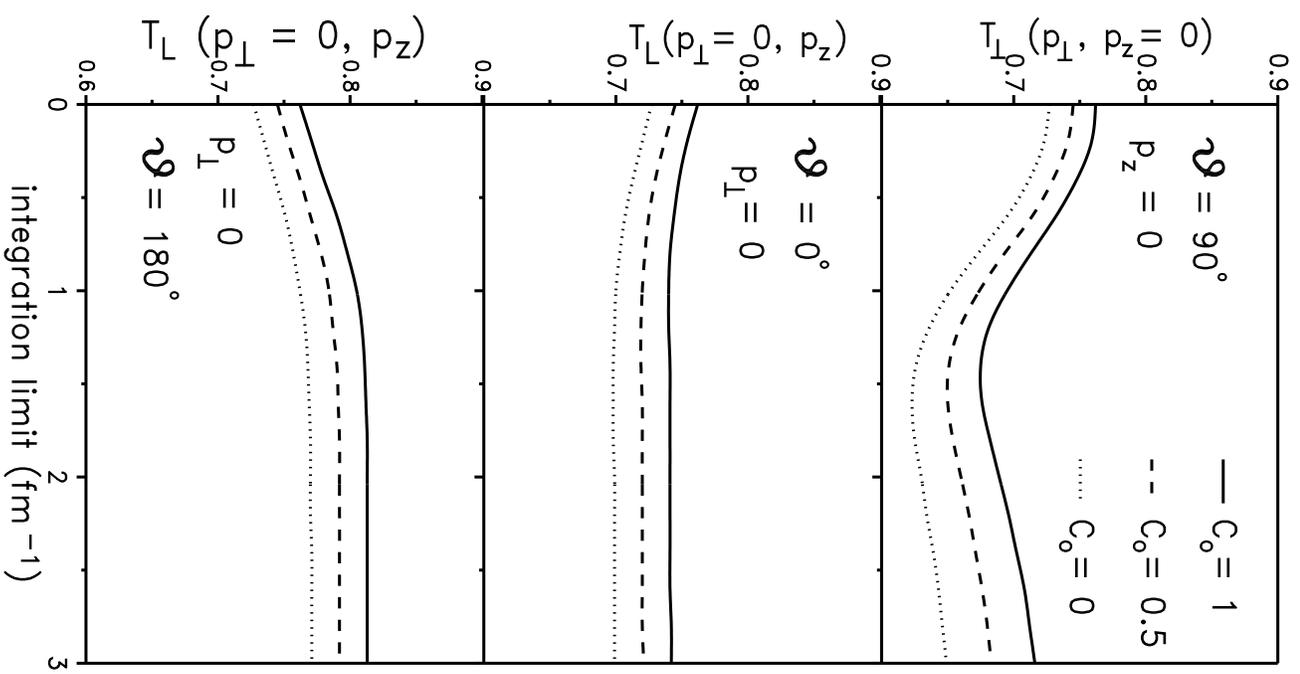
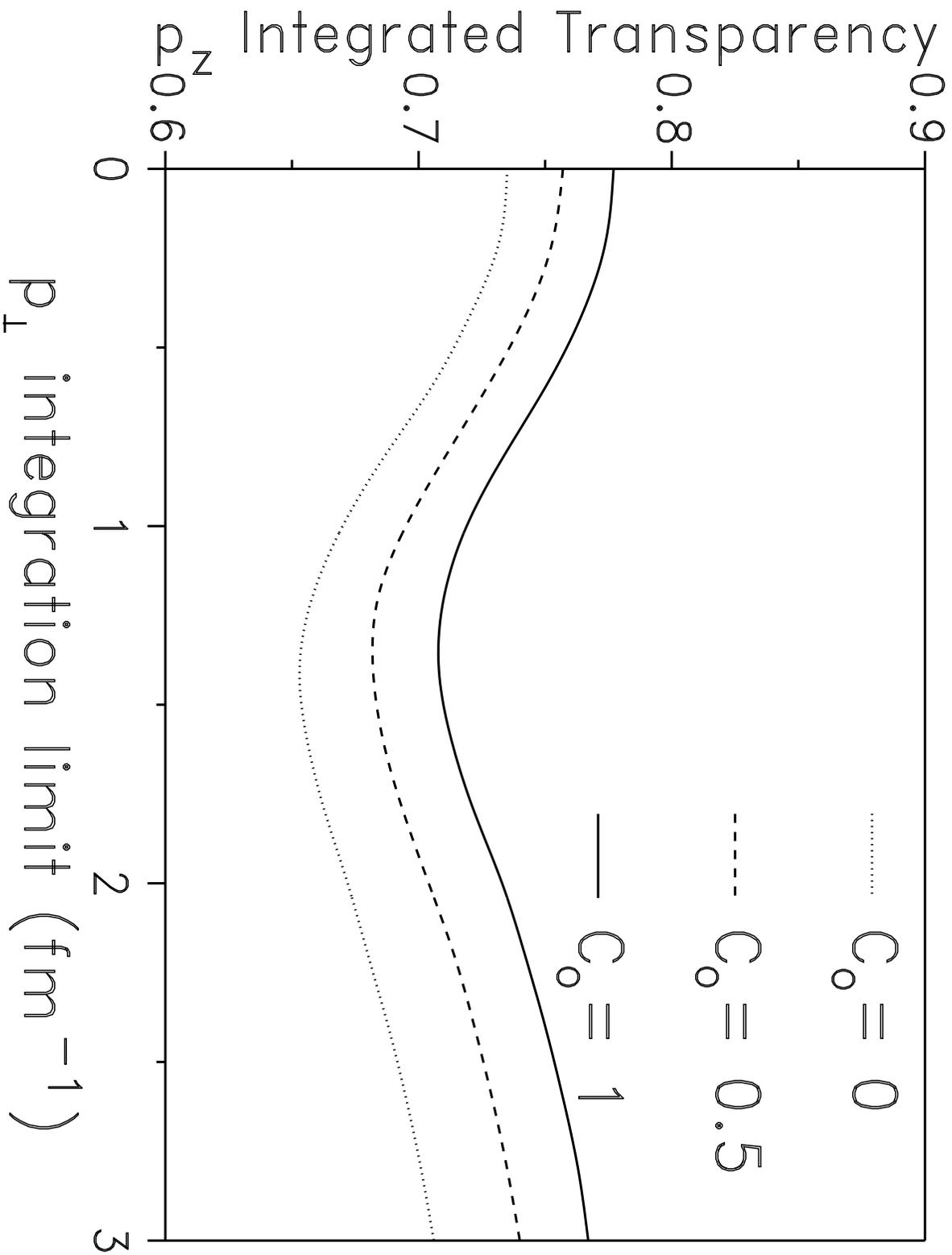
\end{document}